\documentclass[14pt, preprint2, twocolumn]{aastex6}
\usepackage{graphicx, subfigure, amsmath, morefloats, mathtools, amsfonts}

\usepackage{longtable}
\newcommand{\project}[1]{\textsl{#1}}

\newcommand{\apogee}{\project{\textsc{apogee}}}

\newcommand{\kepler}{\project{Kepler}}

\newcommand{\Gaia}{\project{Gaia}}
\newcommand{\Gaiaeso}{\project{Gaia--\textsc{eso}}}
\newcommand{\galah}{\project{\textsc{galah}}}

\newcommand{\documentname}

\newcommand{\teff}{\mbox{$T_{\rm eff}$}}

\newcommand{\sigint}{\mbox{$\rm \sigma_{int}$}}

\newcommand{\logg}{\mbox{$\log g$}}

\newcommand{\msun}{\mbox{$\rm M_{\odot}$}}
\newcommand{\dexgyr}{\mbox{$\rm dex~Gyr^{-1}$}}

\newcommand{\xone}{\ensuremath{x_n}}
\newcommand{\xtwo}{\ensuremath{x_{n'}}}


\begin{document}

\title{{The relationship between age, metallicity, and  abundances for disk stars in a simulated Milky Way galaxy }}

\author{Andreia Carrillo \altaffilmark{1,2,3}}
\author{Melissa K. Ness \altaffilmark{2,4}}
\author{Keith Hawkins \altaffilmark{1}}
\author{Robyn Sanderson \altaffilmark{2,5}}
\author{Kaile Wang \altaffilmark{1}}
\author{Andrew Wetzel \altaffilmark{6}}
\author{Matthew A. Bellardini \altaffilmark{6}}

\altaffiltext{1}{Department of Astronomy, University of Texas at Austin, 2515 Speedway, Stop C1400, Austin, TX 78712-1205, USA}\
\altaffiltext{2}{Center for Computational Astrophysics, Flatiron Institute, Simons Foundation, 162 Fifth Avenue,
New York, NY 10010, USA}\
\altaffiltext{3}{Institute for Computational Cosmology, Department of Physics, Durham University, Durham DH1 3LE, UK}\
\altaffiltext{4}{Department of Astronomy, Columbia University, Pupin Physics Laboratories, New York, NY 10027, USA}\
\altaffiltext{5}{Department of Physics and Astronomy, University of Pennsylvania, 209 S 33rd St., Philadelphia, PA 19104, USA}\
\altaffiltext{6}{Department of Physics and Astronomy, University of California, Davis, CA 95616}\

\email{andreia.carrillo@durham.ac.uk}

\begin{abstract} 
Observations of the Milky Way’s low-$\alpha$ disk show that at fixed metallicity, [Fe/H], several element
abundances, [X/Fe], correlate with age, with unique slopes and small scatters around the age-[X/Fe]
relations. In this study, we turn to simulations to explore the age-[X/Fe] relations for the elements C, N, O, Mg,
Si, S, and Ca that are traced in a FIRE-2 cosmological zoom-in simulation of a Milky-Way like
galaxy, m12i, and understand what physical conditions give rise to the observed age-[X/Fe] trends.
We first explore the distributions of mono-age populations in their birth and current locations, [Fe/H], and [X/Fe], and find evidence for inside-out radial growth for stars with ages $<$ 7 Gyr. We then examine the age-[X/Fe] relations across m12i's disk and find that the direction of the trends agree with observations, apart from C, O, and Ca, with remarkably small intrinsic scatters, \sigint~(0.01 $-$ 0.04 dex). This \sigint~measured in the simulations is also metallicity-dependent, with \sigint~$\approx$ 0.025 dex at [Fe/H] $= -0.25$ dex versus \sigint~$\approx$ 0.015 dex at [Fe/H] $= 0$ dex, and a similar metallicity dependence is seen in the GALAH survey for the elements in common. Additionally, we find that \sigint~is higher in the inner galaxy, where stars are older and formed in less chemically-homogeneous environments.
The age-[X/Fe] relations and the small scatter around them indicate that simulations capture similar chemical enrichment variance as observed in the Milky Way, arising from stars sharing similar element abundances at a given birth place and time. 
\end{abstract}

\section{Introduction}

Galaxy  formation  is  a  violent,  chaotic  process. Small galaxies are stripped of their stars and accreted \citep{searle78} by their larger counterparts, cold  gas is  accreted  through filaments (e.g., \citealt{keres05}), and structures such as spiral arms and bars form and disrupt stellar orbits (e.g., \citealt{roskar08}).  Yet, in some ways,  galaxy  formation  is  also very  orderly. Across populations, we observe that star formation proceeds from the central regions to the outskirts (e.g., \citealt{bezanson09}; \citealt{carrillo20b}) as exhibited by abundance gradients with radius \citep{Nidever2014,Hayden2015, kaplan16, belfiore17,Weinberg2018}, and disk scale heights decrease as galaxies evolve \citep{wisnioski15}. The combination of these orderly and disorderly  processes  gave  rise  to  our very  own  Milky Way. 

With Galactic archaeology, and specifically through using stars as our rosetta stone, we can begin to disentangle the detailed formation history of our Galaxy. This is because stars exhibit atmospheric abundances that generally reflect the chemistry of the gas from which they formed, barring dredge-up processes that bring elements made in the core of stars onto the surface \citep{iben65}. Although a star's orbital and kinematic properties can change as it migrates from where it was born, its chemical fingerprint or inventory is an intrinsic property that remains relatively unchanged throughout its lifetime. With every successive stellar birth and death, or mass loss from stellar winds, the gas from which newer stars will be formed is further enriched. Thereby, stars will have locked within them the chemical fingerprints inherited from their environment, as expressed in the \textit{time and place} of their  formation. 

In the last decade, we have been able to peer into the detailed chemical abundances of different stellar populations in the Milky Way with large spectroscopic surveys. Large multi-object spectroscopic stellar surveys, such as the Apache Point Observatory Galactic Evolution Experiment (\apogee, \citealt{Majewski2017}), the Galactic Archaeology with HERMES (\galah, \citealt{deSilva2015}), the \Gaiaeso~survey \citep{Gilmore2012}, and the Large Sky Area Multi-Object Fibre Spectroscopic Telescope (LAMOST, \citealt{cui12}) have expanded our stellar census of the Galaxy. These surveys have enabled an extensive characterization of the Milky Way’s two-component disk, historically called thin and thick disks due to their different spatial properties, and later on understood in terms of the low-$\alpha$ and high-$\alpha$ disks, respectively \citep{blandhawthorn19}. Earlier work (e.g., \citealt{chiappini97}) have shown that these two populations have different formation timescales, and with these large Milky Way surveys, it has been further solidified that the two components are spatially, kinematically, and chemically distinct, with varying contributions as a function of their location in the Galaxy \citep{Hayden2015}. Many groups have posited different origins for this chemical bimodality including clumpy star formation \citep{Clarke2019}, radial migration \citep{sharma20_RM}, or the presence of a (gas-rich) merger \citep{grand20,lian20,buck20,agertz21}. Following on from this example, chemical abundances, especially for a large number of stars, e.g., N $> \rm 10^{5}$, are key to understanding how the Milky Way built its disk.  While chemistry informs which stars formed together, age ultimately ties these processes to the temporal evolution of the Galaxy.  

There have been many advances in the realm of age derivation for individual stars within large surveys. This has largely been driven by the availability of a set of reference objects with precision ages. The \kepler~mission, in particular \citep{borucki10}, has provided asteroseismic ages for red giant branch stars (RGB) \citep{pins14,Pins2018}. The \Gaia~satellite \citep{Gaia2018} has also measured precision parallaxes, allowing age derivation from isochrones around the main sequence turn off (MSTO). Leveraging the mass-dependent, dredge-up-induced [C/N] ratio at the stellar surface \citep{Masseron2015,martig16}, data-driven approaches that use high precision reference ages have enabled ages to be determined from large samples of spectra (e.g. \citealt{ness16_ages,ho17_ages}).

Ages have provided the means to understand nucleosynthesis over time, and link the chemical evolution to the assembly history of a galaxy. There is a known age-[Fe/H] relationship for the Milky Way disk, wherein stars decrease in [Fe/H] with older ages (e.g., \citealt{twarog80}; \citealt{soubiran08}). However, whether there is a clear trend is debated in observational studies: previous works \citep{Haywood2013,bergemann14,rebassamanserga16} found that there is substantial scatter in the relation, with \citet{nissen15} finding this lack of correlation existing over a large age interval (e.g., 8 Gyr). This same study, however, found that there is a tight correlation between the age and [X/Fe] of stars. It is important to further investigate these age-individual abundance, or age-[X/Fe], relations at a given [Fe/H] in the disk because the tightness of these relations 
illustrate that (1) by inverting the age-[X/Fe] relation, abundances can serve as chemical clocks providing ages for stars, and that (2) [Fe/H] and age capture majority of the crucial information about a star e.g., [X/Fe], orbits, and birth location \citep{bedell18,ness19, jofre20,hayden20,espinozarojas21,casamiquela21,sharma22,ratcliffe22}. Additionally, comparisons between the slope in the [Fe/H]-[X/Fe] relation and the slope in the age-[X/Fe] relation appear to group elements accordingly into three distinct nucleosynthetic sites: core-collapse supernova (SNe II), white dwarf explosion (SNIa), and stellar winds \citep{sharma22}. Furthermore, the slope in the age-[X/Fe] relation is sensitive to the location across the [Mg/Fe] vs [Fe/H] plane \citep{lu21, horta2021}, but the scatter around these relations is consistently small. Both the high resolution studies at [Fe/H] = 0 dex from \citet{nissen15} and \citet{bedell18} as well as the larger samples exploring a wider range of metallicities \citep{ness19,sharma22} have shown that the intrinsic scatter of stars around their age-[X/Fe] relations is on the order of $<$ 0.01-0.04 dex. The small intrinsic scatter in the age-[X/Fe] relations is also fairly insensitive to the \textit{current} location of the stars in the disk \citep{ness19,lu21}, which is a natural consequence of radial redistribution or migration, whereby stars at any given radius, R, are from a wide range of initial radii \citep{Frankel2018,frankel19}.

The growing literature on the age-[X/Fe] relations in observations could constrain the nucleosynthetic processes that produce the elements in the universe and the star formation history and interstellar medium conditions that gave rise to the observed age-[X/Fe] relations and scatters in the Milky Way. 
This is because stars not only remember their birth sites through their detailed chemistry, but for an established disk, these birth sites, at a given age, can be ordered in radius by sorting in metallicity. Likewise, for a given metallicity, sorting in age probes systematically different birth radii. The observed low intrinsic scatter around the age-[X/Fe] trends at fixed [Fe/H] indeed corroborate this picture, wherein stars are formed with element abundances governed by their birth radii and time of formation.  

It is therefore timely that we understand what gives rise to this relationship in a self-consistent manner. That is, investigating this relationship where we know the galaxy formation history and we know the nucleosynthesic processes that together bring about the resulting chemistry. Here, we utilize cosmological simulations as a means to examine the age-[X/Fe] relations across the disk. Within these simulations, we have access to the ages and chemistry of stars with a direct knowledge of birth properties while unhindered by observational uncertainties. Critically, recent cosmological zoom-in simulations (e.g. \citealt{wetzel16}, \citealt{grand16}, \citealt{buck2019_nihao}, \citealt{wetzel22}) now implement physically motivated processes (e.g. sub-grid turbulent metal diffusion, stellar feedback, chemical enrichment from SNe II, SNIa, and stellar winds) at high resolutions that can distinguish individual star forming regions. 

Galactic archaeology studies with simulated Milky Way-like galaxies have proven themselves instructive in demonstrating physical processes that can give rise to observed trends. For example, cosmological simulations show that gas-rich mergers can give rise to the low-$\alpha$ sequence seen in the Milky Way disk \citep{buck20,agertz21}. \citet{yu21} used the Feedback in Realistic Environments 2 (FIRE-2) simulations \citep{hopkins18,wetzel22}, to demonstrate that the thick disk formation phase in these Milky Way-like galaxies coincides with the star formation changing from a bursty mode to a steady mode. \citet{nikakhtar21} similarly used FIRE-2 Milky Way-like galaxies and explored the different families of stars in the solar neighborhood, determined through a Gaussian mixture modeling of kinematics and [Fe/H]. They found simulated analogs to the real and observed stellar population distinctions in the Milky Way solar neighborhood, and subsequently related these to having different origins. Specifically, they found a thin disk component that is young, a halo component consistent with early and massive accretion events, and three thick disk components with heated orbits due to satellite interactions. \citet{bellardini21,bellardini22} also looked at [Fe/H] and abundance gradients in FIRE-2 Milky Way-like galaxies, where they found that the [Fe/H] and abundances change from being dominated by azimuthal variations to being dominated by radial variations at lookback times 7--7.5 Gyr ago.  This age broadly coincides with an earlier study by \citet{ma17} as the transition from a chaotic, bursty mode of star formation to a calmer, stable disk in the FIRE-1 galaxy m12i.  As illustrated by these studies, the star formation history, gas inflow and outflow, nucleosynthetic processes, and environments that these galaxies live in are known, and therefore aid in understanding the possible origins of observed stellar population characteristics and interpret chemo-dynamical signatures in the Milky Way. 

Therefore, in this study, we similarly take advantage of a cosmological zoom-in simulation of a Milky Way-like galaxy to explore and establish the relationship of the age, metallicity, and individual element abundances of stars.  We use the Ananke Gaia synthetic survey \citep{sanderson20} to understand the observed tight age-[X/Fe] trend at a fixed [Fe/H]. Specifically, we aim to investigate this trend at different [Fe/H] and locations in the disk,  and to show where the simulations reproduce this observed relationship, where it breaks down, and how this gap could be bridged. Ultimately, we aim to be able to put into context the physical processes in the simulated galaxy that give rise to the age-[X/Fe] trend it exhibits. In Section \ref{sec:sim_data} we describe the simulation data we use for this work and the galaxy m12i. In Section \ref{sec:obsdata} we discuss the observational data we use for comparison. In Section \ref{sec:kdes} we explore the abundance and location distributions of stars with similar ages. In Section \ref{sec:abund_age_trends} we show the age-[X/Fe] trends for the various elements that are tracked in the simulations at different [Fe/H] and galactocentric radii. In Section \ref{sec:obscomparison} we compare the intrinsic scatter in the age-[X/Fe] trends between observations and simulations. Lastly, in Section \ref{sec:discussion_cca} we discuss the implications of our results and provide a summary of this work. We note that for the majority \footnote{In Section \ref{sec:kdes}, we analysed one aspect of birth location of stars, i.e., their distributions with age, which is not public data.} of the analysis, we used the publicly available information in the Ananke Gaia synthetic survey.

\section{Simulation Data}
\label{sec:sim_data}

\subsection{Ananke: Gaia Synthetic Surveys}
\label{sec:ananke}

We take advantage of the Gaia synthetic surveys produced with the Ananke framework \citep{sanderson20} which generates synthetic phase-space surveys from baryonic simulations. These synthetic surveys are based on the \textit{Latte} suite of zoom-in simulations of Milky Way-like galaxies \citep{wetzel16} utilizing FIRE-2 physics \citep{hopkins18}, that features state-of-the-art implementations of hydrodynamics, radiative cooling and heating, star formation, and stellar feedback.

In this simulation suite, stars are formed from self-gravitating molecular gas that is self-shielding (following \citealt{krumholz11}), dense (n $>$ 1000 $\rm cm^{-3}$), cold (T$<$ 10,000 K), and Jeans-unstable---prescriptions that produce clustered stellar populations naturally. A star particle is essentially a single stellar population with an initial mass of 7070~\msun~but with mass loss from stellar winds, it reduces to a typical mass of $\sim$5000~\msun~ at z = 0. 

The stellar evolution of star particles is produced from STARBURST99 \citep{leitherer99} with  a \citet{kroupa01} initial mass function (IMF). Chemical enrichment occurs through three main nucleosynthetic sources: SNe II, SNIa, and OB/asymptotic giant branch (AGB) stellar winds. The yields for SNe II are from \citet{nomoto06}, for SNIa from \citet{iwamoto99}, and OB/AGB winds from \citet{wiersma09} compiling values from \citet{vdhoek97}, \citet{marigo01}, and \citet{izzard04}. SNIa rates are taken from \citet{mannucci06} including both prompt and delayed populations. We note that the yields in the simulations are IMF-averaged \citep{hopkins18}. These stellar and chemical evolution prescriptions (listed in Appendix \ref{sec:appendix_yields}) affect the present-day ``observed" abundances in the synthetic survey which is central to our analysis. 

Sub-resolution gas metal diffusion is included, which leads to abundance distributions that better match observations \citep{su17,hopkins18,escala18}. For a more detailed exploration of how the diffusion coefficient affects abundance scatters in Milky Way-like galaxies, see \citet{bellardini21}, Appendix C. 

The simulations track the total mass of each element for each star particle, but observationally, the abundance measurement is in terms of [X/Fe], where  [X/Fe] = [X/H] - [Fe/H] and the bracket,``[ ]", notation denotes relative abundance with respect to Solar abundance values. To put the abundances from the simulations in the same scale, \citet{asplund09} Solar values were used to calculate
\begin{equation}
    \rm [X/H] = \frac{m_{X}/m_{X,\odot}}{m_{H}/m_{H,\odot}},
\end{equation}
where X is the element, $\rm m_{X}$ is the mass of the element in the star particle, and $\rm m_{X,\odot}$ is the Solar value for that element.

\begin{figure*}
\centering
\includegraphics[width=0.8\textwidth]{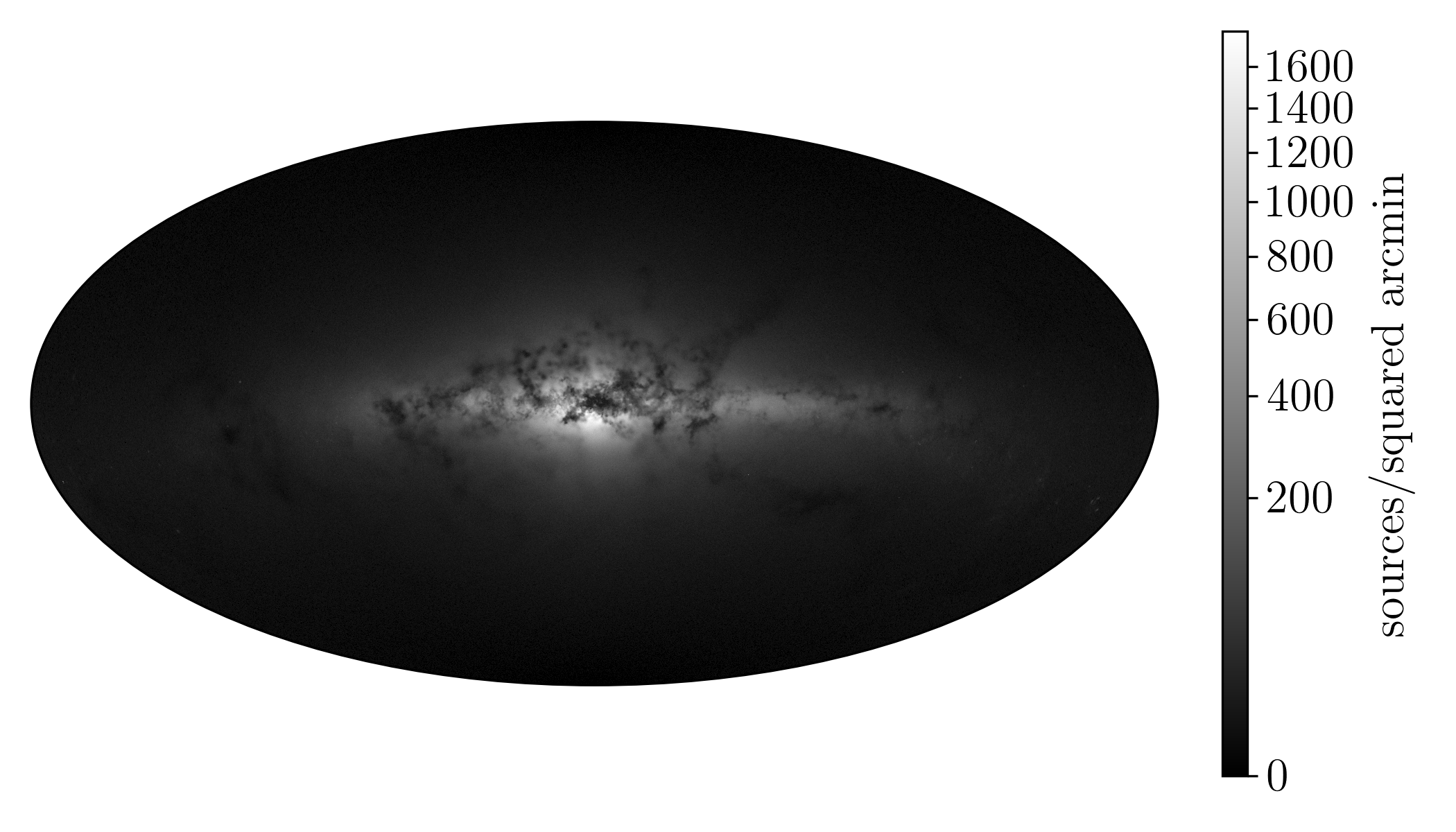} 
\caption{\textit{A picture of m12i}. Aitoff projection of the star counts in m12i viewed from LSR2. The galaxy shows structure similar to the Milky Way i.e. a thin disk, a more diffuse thick disk, a central bulge-like concentration of stars, and patchiness due to dust extinction.} 
\label{fig:galacticmap}
\end{figure*}

Three galaxies were taken from \textit{Latte}, labeled m12f, m12i, and m12m, for the generation of the \textit{Gaia} synthetic surveys viewed from three different ``Solar" view points per galaxy (resulting in nine synthetic surveys in total), where a ``Solar" view point (Local Standard of Rest, LSR) is defined as 8.2 kpc from the galactic center. These Milky Way-like galaxies were selected based \textit{only} on their total mass ($\rm M_{200m}$\footnote{$\rm M_{200m}$ is a proxy for virial mass in simulations \citep{white01} where $\rm 200m$ is 200 $\times$ the matter density} = 1-2 $\rm \times 10^{12} M_{\odot}$) and environment (isolated, i.e., has no nearby dark matter halo with similar mass up to 5 $\rm \times R_{200}$\footnote{$R_{200}$ is a proxy for virial radius \citep{white01}}) and therefore have a variety of morphologies, satellite populations, and star formation histories. Each catalog has self-consistent dust extinction implemented, based on scaling of the gas density of the simulated galaxies. For a proof of concept in exploring age-[X/Fe] trends in simulations, and for simplicity, we focus our analysis on one viewpoint (LSR 2) of the galaxy m12i which is also publicly available \footnote{\url{https://girder.hub.yt/##collection/5b0427b2e9914800018237da}}. The Galactic map of m12i is shown in Figure \ref{fig:galacticmap}.
 
For the synthetic surveys, the single stellar populations are broken into individual synthetic stars that are sampled similarly from a \citet{kroupa01} IMF and mapped to a grid of isochrones in age and [Fe/H] over a set of model isochrones (PARSEC; \citealt{bressan12},\citealt{marigo13},\citealt{marigo17}). We note that the isochrones used for generating the synthetic surveys are different from the isochrones used for the stellar evolution in the simulations. Specifically, they differ in their treatment of high-mass star evolution where the simulations use the Geneva tracks \citep{lejenune97} that have enhanced mass loss from rotation while PARSEC includes thermally-pulsating AGB stars. However, \citet{sanderson20} prescribed mass-independent mass-loss rate in sampling the number of stars from a star particle, alleviating some of the differences introduced from the mismatch in isochrones.

These individual stars then inherit the age, [Fe/H], and [X/H] of the parent star particle, while their 6D phase-space information is sampled over a one-dimensional kernel in each position and velocity axis centered on the parent star particle. We use the true values for the ``observed" properties (e.g. distance, age, and chemistry) of stars in the synthetic survey because the essence of our analysis is to understand the \textit{intrinsic} scatter in the age-[X/Fe] trend. However, we selected the stars from the synthetic survey to mimic the constraints in observations which we ultimately draw comparisons to. 

\subsection{Selection of stars for analysis}
\label{sec:sample_selection}

\begin{figure*}
\centering
\includegraphics[width=0.75\textwidth]{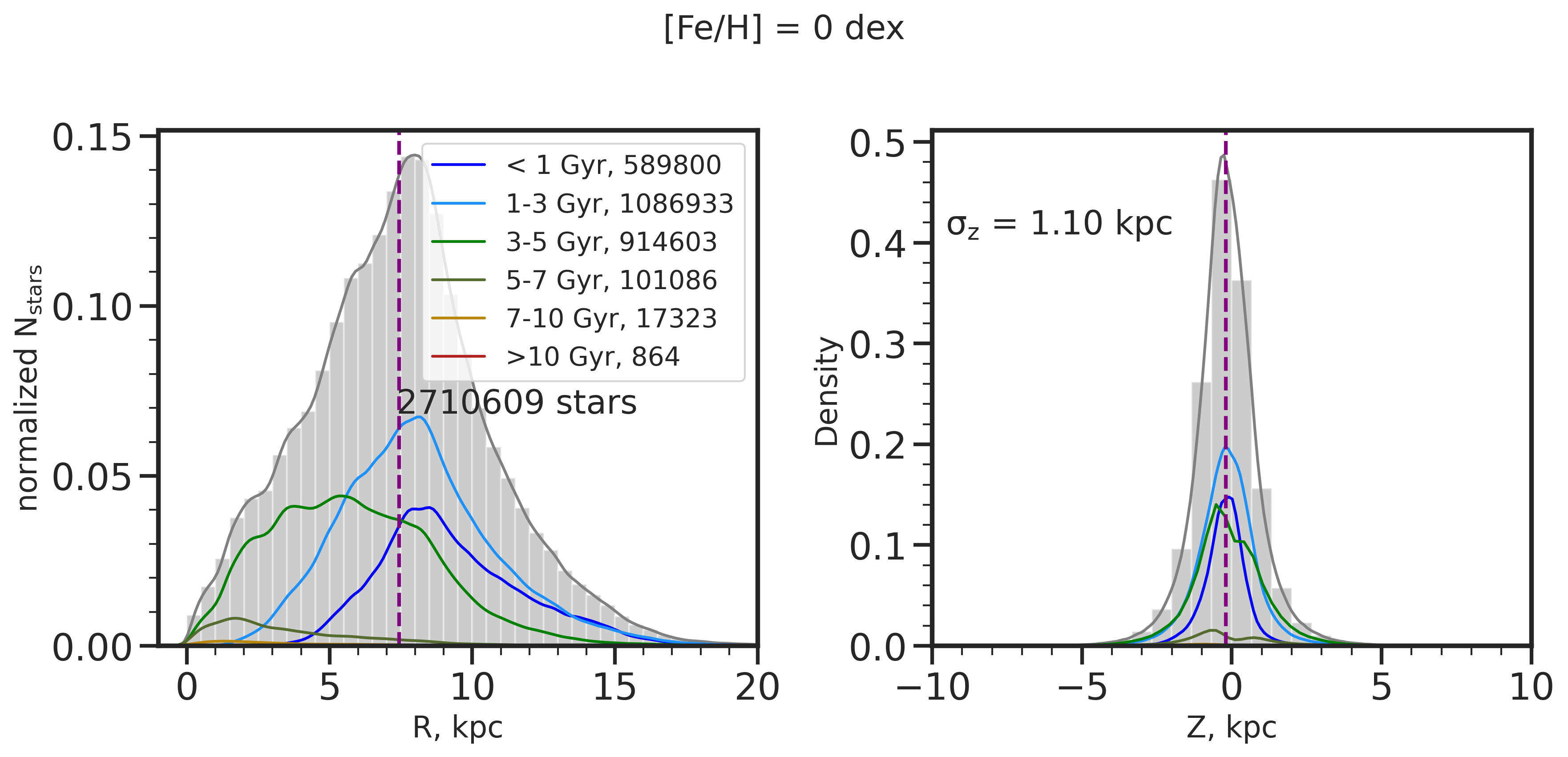}
\includegraphics[width=0.75\textwidth]{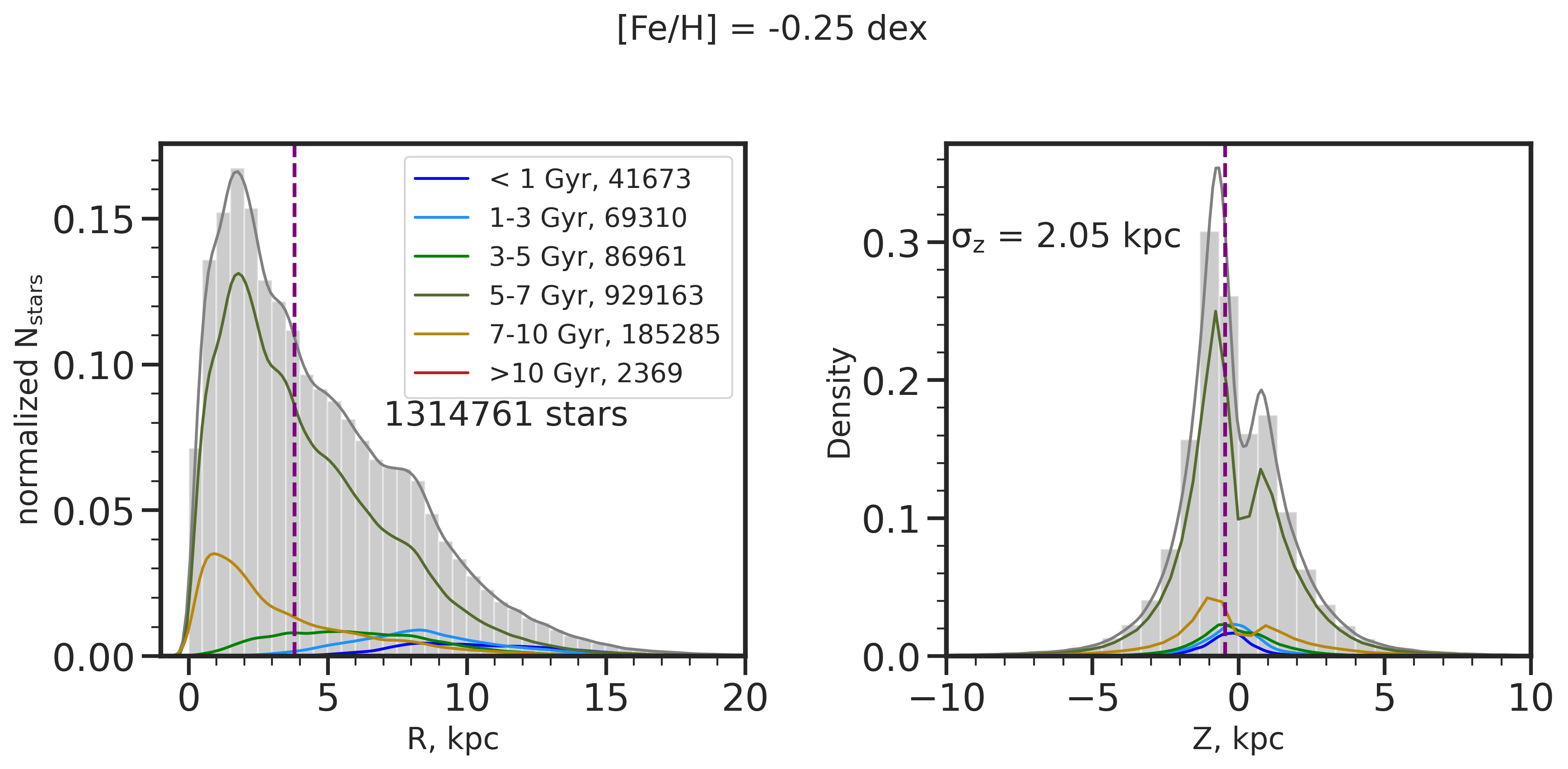} 
\includegraphics[width=0.75\textwidth]{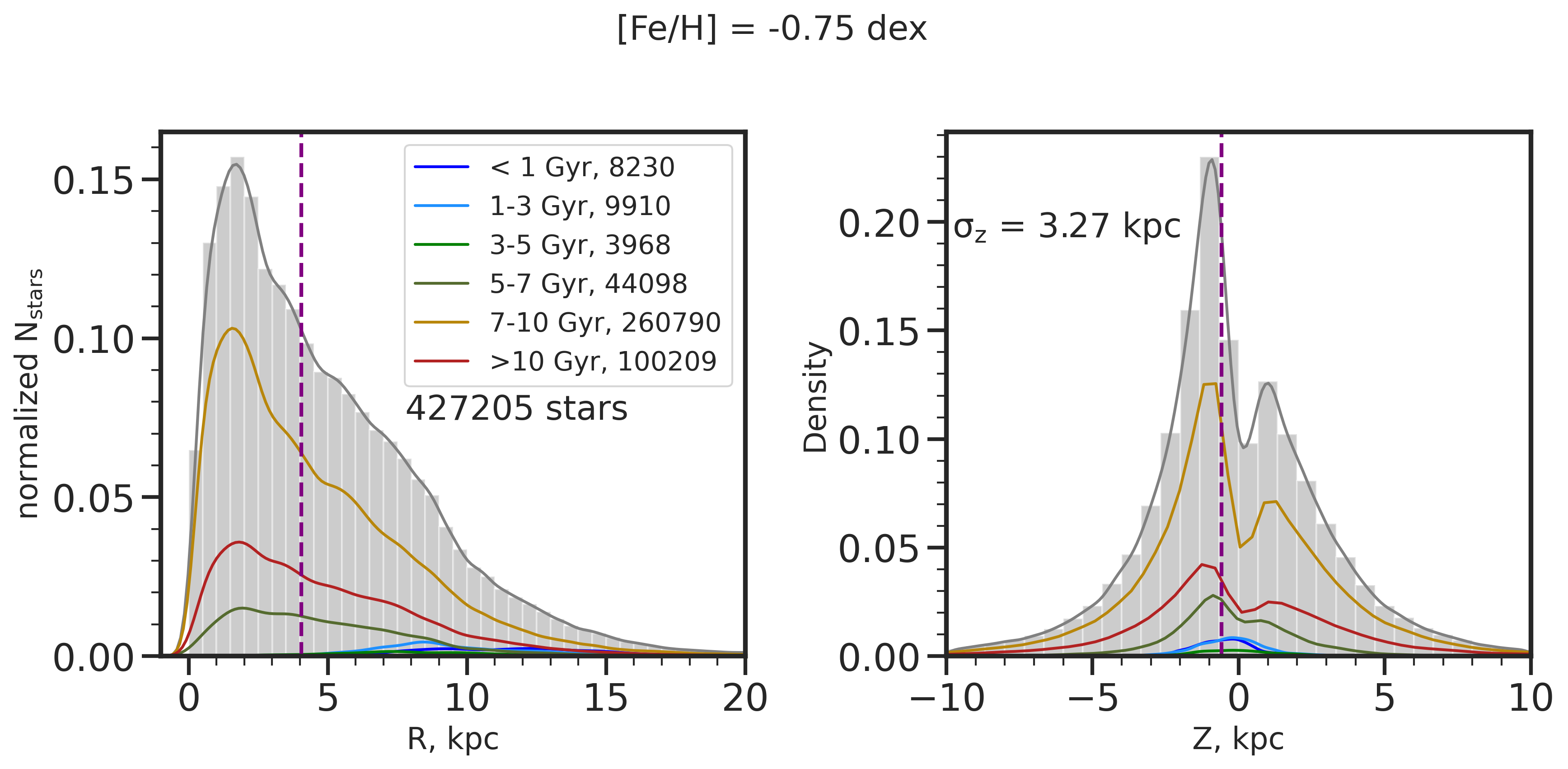} 
\caption{\textit{R and Z distributions at different [Fe/H] and ages}. We select three different samples of stars: one at solar metallicity ([Fe/H] = 0 dex, top), and two at low metallicities ([Fe/H] = -0.25 dex, middle and [Fe/H] = -0.75 dex, bottom) and plot their current R (left) and Z (right) distributions. For each [Fe/H] bin that has a size of $\pm$0.025 dex, we mark the distribution of stars at different ages i.e. $<$1 Gyr (light blue), 1-3 Gyr (dark blue), 3-5 Gyr (green), 5-7 Gyr (olive), 7-10 Gyr (gold), and $>$10 Gyr (brick red). We also mark the median R and Z at these metallicities (dashed purple line). The R distributions move from the inner to the outer regions while the Z distributions change from broad to narrow as we go from older to younger stars. } 
\label{fig:rnowhist}
\end{figure*}

In this work, we aim to compare the age-[X/Fe] relations for the thin disk stars in the simulated Milky Way to those measured in the observed Milky Way low-$\alpha$ disk (e.g. \citealt{bedell18}, \citealt{ness19}), that have in-situ origins and have been the subject of most age-[X/Fe] studies. With this in mind, \textbf{we have  constrained our analysis of stars in the simulations to those that (1) have R$<$25 kpc, (2) [Fe/H] $\geq$ -0.25 dex, and (3) $\mid Z\mid < 500$ pc.} We do this selection to obtain predominantly in-situ stars in the current day spatial thin disk, though we note that there is no explicit removal of ex-situ stars. We explain these criteria in detail below. 
We note that for ease of processing, we utilize the subsampled independent survey that is $1/100$ the size of the full synthetic survey for m12i at LSR2. 

The first criterion (1) restricts to stars within 25 kpc of the galactic center. This is motivated by Figure 3 of \citet{sanderson18}, which shows the current locations of stars in the FIRE simulated galaxies colored by their formation locations. For m12i, these authors found that at R$\sim$20 kpc, there is a sharp transition between stars that formed in-situ and stars that formed at greater distances, i.e. R $>$30 kpc. Azimuthally, this transition radius hovers around the R = 20 kpc value, though  sometimes exceeds this range. We therefore opted to use R = 25 kpc as our radius cut to take into account the upper bound of this transition region. Interestingly, \citet{sanderson18} also found that this transition region between mostly in-situ vs mostly accreted stars is also traced by a transition in [Fe/H], which is a promising diagnostic for observational studies. This is also similar to the selection made in \citet{bellardini22} in defining the in-situ component of Milky Way-like galaxies in FIRE-2. 

As a sanity check for this selection, we investigate how the spatial distributions, specifically in current (cylindrical) Radius, R, and current height from the disk plane, Z, of our resulting sample change as a function of age and metallicity. This is also motivated from observations \citep{ness19,lu21} that the tight relationship between [X/Fe] and age at fixed [Fe/H] is due to stars reflecting the chemistry of their location; therefore, we expect these distributions to be distinct at different ages and [Fe/H]. In Figure \ref{fig:rnowhist}, we show the R distributions on the left and Z distributions on the right for stars at different [Fe/H]: solar at 0 dex (top), -0.25 dex (middle), and -0.75 dex (bottom), with a bin size of [Fe/H] $\pm$0.025 dex, and within R $<$ 25 kpc of the galactic center. We have explored stellar populations in other [Fe/H] bins (from [Fe/H] = +0.25 dex to -2.5 dex) as well, but focus on these three, 
which are sufficient to succinctly portray the differences in the R and Z distributions at high and low [Fe/H]. Each R and Z distribution is broken down into the separate contributions from different stellar ages, i.e., $<$1 Gyr (light blue), 1-3 Gyr (dark blue), 3-5 Gyr (green), 5-7 Gyr (olive), 7-10 Gyr (gold), and $>$10 Gyr (brick red). We also mark the median R and Z values with the purple dashed line. Though our first criterion makes a cut at R$<$25 kpc to select the predominantly in-situ population, not many stars lie in that region. In order to better compare the R trends at different ages, we therefore zoom in and only show the distributions up to R = 20 kpc.  

At solar metallicity, the R distribution peaks at R = 8 kpc, slightly coinciding with the median R marked by the purple dashed line at R = 7.5 kpc. Meanwhile the R distributions for the lower metallicity samples (i.e. [Fe/H] = -0.25 dex and -0.75 dex) peak close to, but not quite at the center of the galaxy at R = 2 kpc, in line with the results from \citet{elbadry18} for the oldest stars in the FIRE-2 simulations. The median R for both low metallicity bins is higher than the peaks in their distributions and lies at R = 4 kpc, as both distributions have long tails towards larger R. For the Z distributions, the trend at solar metallicity resembles a normal distribution with mean at Z = 0. At lower metallicities, however, the distributions deviate from a Gaussian and exhibit two peaks, below and above the plane of the disk. The dispersion in Z also increases from high to low [Fe/H], as indicated on the upper left corner of each subplot in Figure \ref{fig:rnowhist}, from $\sigma_Z$ = 1.10 kpc at [Fe/H] = 0, to $\sigma_Z$ =  2.05 kpc at [Fe/H] = $-0.25$ dex, and to $\sigma_Z$ = 3.27 kpc at [Fe/H] = $-0.75$ dex. This trend of increasing Z dispersion with increasing age and decreasing metallicity exists in the Milky Way as well \citep{meusinger91}, and could be explained by stars being born in a thicker disk at early times \citep{wisnioski15,ma17,bird21,yu21}, or by being heated kinematically through interactions with giant molecular clouds, spiral arms, and bar \citep{aumer16} and with satellites \citep{hopkins08,villalobos08,villalobos09,sales09,laporte19}. On the other hand, the double-peak in the Z distribution and the asymmetry about Z = 0 is due to the fact that it is harder to see stars through the disk because of the dust and that older stars have larger scale heights. 

In general, we find that the solar, [Fe/H] = 0 dex bin is dominated by stars with ages $<$5 Gyr, at [Fe/H] = -0.25 dex by stars that are $5-7$ Gyr old, and at [Fe/H] = -0.75 dex by stars that are $>$7 Gyr old. As expected, younger stars dominate at high [Fe/H] and older stars dominate at lower [Fe/H]. These stellar populations' respective R distributions therefore also dominate the total R distribution for each [Fe/H] bin. For example, although the [Fe/H] = -0.25 dex bin is comprised of stars of different ages,  the dominant stellar age is 5-7 Gyr (olive green line); therefore, the total R distribution (gray), across all ages, consequently follows the R distribution of the $5-7$ Gyr stellar population. We note, however, that within the same [Fe/H] bin, these R (and Z) distributions at different ages highly overlap. 

At [Fe/H] = 0, the R distribution peak at higher R values for younger stars than older stars. This is less evident for the lower [Fe/H] bins, where there are fewer younger stars. As seen for the [Fe/H] = $-0.25$ dex sample, the peak of the $5-7$ Gyr distribution has R value very close to that of the $7-10$ Gyr distribution. At the lowest [Fe/H] bin, the R where the distributions peak for stars with ages 5-7 Gyr, 7-10 Gyr, and $>$10 Gyr all coincide at $\sim$2 kpc, showing no sequence in R with the dominant ages. \textbf{Therefore, we primarily observe evidence of an inside-out radial formation for younger and higher metallicity stars, i.e. ages $<$ 7 Gyr and [Fe/H] $>$ -0.25 dex}. That is, only with these cuts do we isolate an inside-out forming disk with a clear current day age gradient at a fixed [Fe/H]. This age is broadly consistent with the transition age found by \citet{yu21} when FIRE-2 galaxies change from a bursty to smooth star formation mode. This is also the transition age that \citet{bellardini22} found for their sample of FIRE-2 Milky Way-like galaxies, when the stellar abundances changed from being dominated by azimuthal variations to radial variations. All Z distributions show increasing dispersion with increasing age at a given [Fe/H]. For the [Fe/H] = 0 dex sample, the Z distributions for stars with ages $<$1 Gyr and 1-3 Gyr are Gaussian, but the distributions for older stars deviate from this behavior. This is more obvious at [Fe/H] = -0.25 dex and [Fe/H] = $-0.75$ dex, which are dominated by older stars that show this double peak more, as the effect of not being able to ``see" through the disk is more significantly felt. These R and Z trends that we find in the simulations are in congruence with the \textit{inside-out and upside-down} scenario for the observed Milky Way stellar age-velocity relation as shown by \citet{bird21}, and is further support for the Milky Way-like nature of this simulation and its earlier version in FIRE-1 (e.g., \citealt{ma17}).
 
With this exploration and these trends in mind, we therefore apply a second criterion in investigating the age-[X/Fe] relations (see Section \ref{sec:abund_age_trends}) and (2) only select stars with [Fe/H] $>$ -0.25 dex to ensure that we are probing the regime in metallicity in which chemical evolution has been temporally smooth, and age and [Fe/H] are a link to a star's location.

Lastly, (3) we spatially constrain our analysis to the simulation's thin disk,  which is the focus of most age-[X/Fe] trend studies in observations (e.g. \citealt{nissen15}, \citealt{bedell18}, \citealt{ness19}). We selected stars with Z within $\pm$500 pc of the disk plane motivated by the measured scale height of the thin disk for m12i \citep{sanderson20} and by the Z distributions of the radially-distinct mono-age, mono-abundance populations shown in Figure \ref{fig:rnowhist}.

As a caveat, we note that this comparison is not entirely consistent because in observational studies, the thin disk selection is usually done in the [$\alpha$/Fe] vs [Fe/H] plane, while in most simulations and in this study, the criteria for the thin disk is in approximation a spatial selection. However, for the purpose of our analysis, this means we compare the stars in the disk with small vertical oscillations around the plane, in both simulation and data.

\section{Observational Data}
\label{sec:obsdata}

\begin{figure}
\centering
\includegraphics[width=0.45\textwidth]{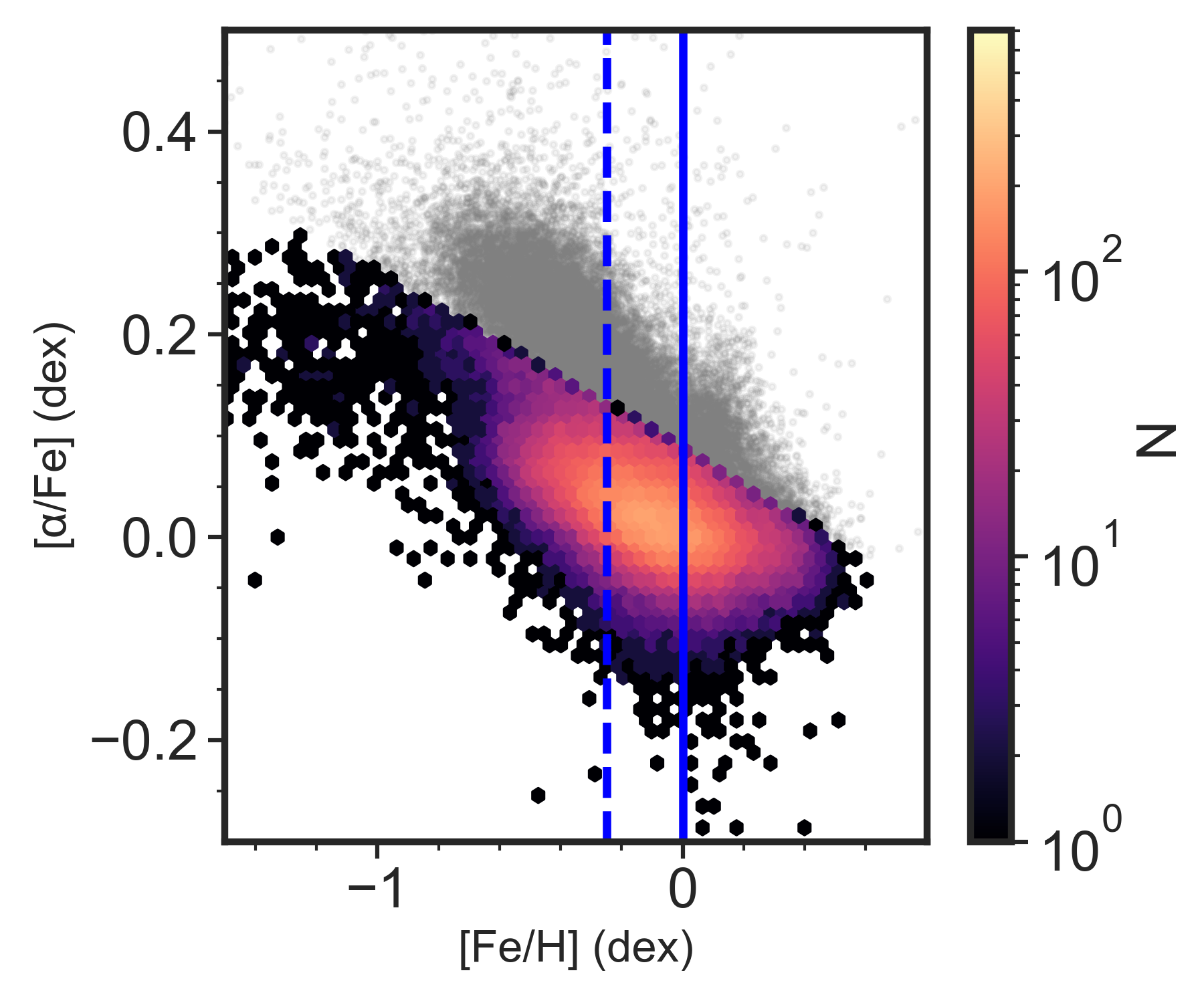} 
\caption{[$\alpha$/Fe] vs [Fe/H] from \galah~DR3. We focus our comparison to the low-$\alpha$ disk, shown as the magma density plot. [Fe/H] = 0 dex is marked with a solid vertical line and [Fe/H] = -0.25 dex is marked with a dashed vertical line. We investigate these [Fe/H] regimes for the discussion in Section \ref{sec:obscomparison}} 
\label{fig:galah}
\end{figure}

We briefly introduce the three sets of observational data that we compare to regarding the [X/Fe] distributions, age-[X/Fe] relations, and the intrinsic scatter, \sigint, around these trends in the simulations. 

First, we use the \galah~survey data release 3 (DR3, \citealt{buder21}). This has an associated value added catalog from \citet{sharma22} that contains ages from the BSTEP code \citep{sharma18} which makes use of stellar isochrones. In total, \galah~DR3 has abundances for 30 elements in 588,571 nearby stars in the Galactic disk. We derived a sample from this data set consisting of 102,785 stars having applied the following quality cuts: signal-to-noise ratio $>$40, flag\_sp=0, flag\_fe\_h=0, and $4000 < \teff < 7000$ K. We further apply a cut in [$\alpha$/Fe] vs [Fe/H] shown in Figure \ref{fig:galah} and as similarly done in previous observational studies (e.g., \citealt{ness19}; \citealt{blandhawthorn19}), to focus on just the low-$\alpha$ disk, leaving us with 81,230 stars. We have the elements C, O, Mg, Si, and Ca in common with the \galah~data and compare their [X/Fe] distributions for mono-age populations with those from the simulations (see Section \ref{sec:kdes}). The study by \citet{sharma22} also provide comparison with age-[X/Fe] relations and the \sigint~around them at varying [Fe/H]. 

Second, we compare to the age-[X/Fe] relations from the set of 79 solar twin stars from \citet{bedell18}, where the ages were derived from the Yonsei-Yale isochrones \citep{demarque04}. This sample of solar twin stars generally have ages $<$8 Gyr. The authors report measurements for 30 element abundances, where they have C, O, Mg, Si, S, and Ca in common with this study. With regards to the [X/Fe]-age trends, the fits were derived through minimizing the orthogonal distance between the data and the model, weighted by the observational uncertainties \citep{hogg10}. 

Lastly, we use the measurements from the set of $\approx$ 15,000 red clump field stars from \citet{ness19} in the low-$\alpha$ disk, where the authors utilize the \apogee~survey's 14th data release \citep{Majewski2017}. This work looked into the age-[X/Fe] relations in 19 different elements, wherein C, N, O, Mg, Si, S, and Ca are similarly measured as in the simulations. The ages were derived through data-driven modeling with \textit{The Cannon} \citep{Ness2015} using the APOKASC catalogue \citep{Pins2018} as a training set which contains ages derived from asteroseismology. This sample of giant stars have ages $<$10 Gyr. In deriving the intrinsic scatter in the [X/Fe]-age trends in their study, the effect of the age uncertainties were taken into account via Monte Carlo sampling. The error in age can shift stars horizontally in the age-[X/Fe] trend. For flat trends, accounting for the age error has minimal effect but for steep trends, this essentially flattens the relationship. This comparison observational work found that the effect of the age uncertainties changes the intrinsic scatter by $<$0.01 dex and was therefore negligible. Likewise, the abundance error was accounted for in deriving the \sigint~as the measured scatter from the best fit line will be a combination of the \sigint~and the error in the abundance measurement, $\rm \sigma_{abund}^2$. That is: $\rm \sigma_{total}^2 = \sigma_{abund}^2 + \sigma_{int}^2$.  

\section{The spatial distribution of stellar chemical abundances at birth}
\label{sec:kdes}
\begin{figure}
\centering
\includegraphics[width=0.45\textwidth]{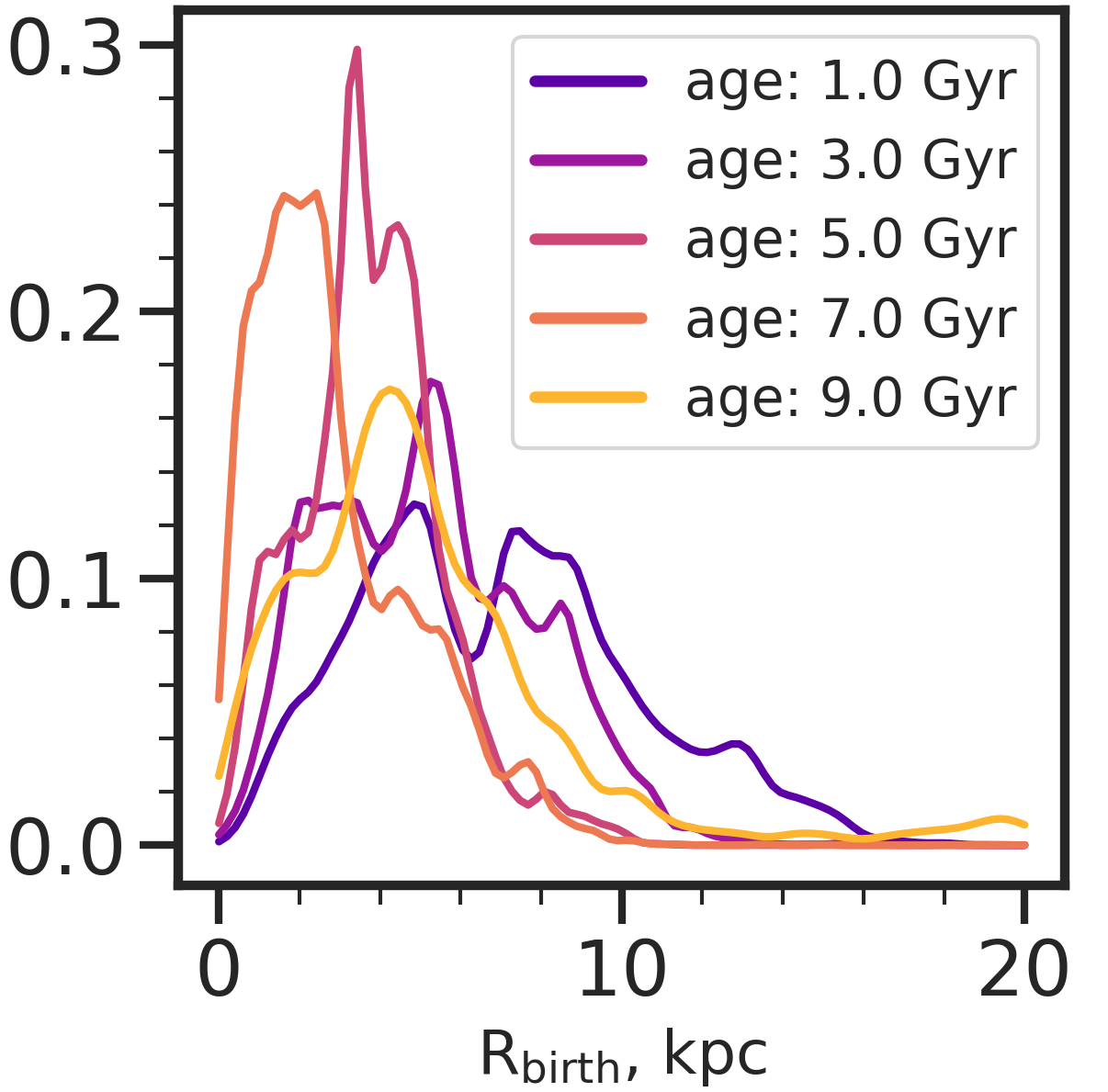} 
\caption{\textit{Birth radius distributions for stars of different ages}. Kernel density estimations (KDE) for $\rm R_{birth}$ 
of stars within age bins centered at 1 Gyr (purple) to 9 Gyr (yellow) in steps of 2 Gyr, prior to the [Fe/H] cut. The KDEs are normalized to the number of stars per age group. 
Stars of different ages trace distributions that, though highly overlapping, show distinct birth annuli. However, the 9 Gyr stellar population does not follow this trend due to a more chaotic mode of star formation at this point in the m12i's history.   } 
\label{fig:R_kdes}
\end{figure}

\begin{figure*}
\centering
\includegraphics[width=0.9\textwidth]{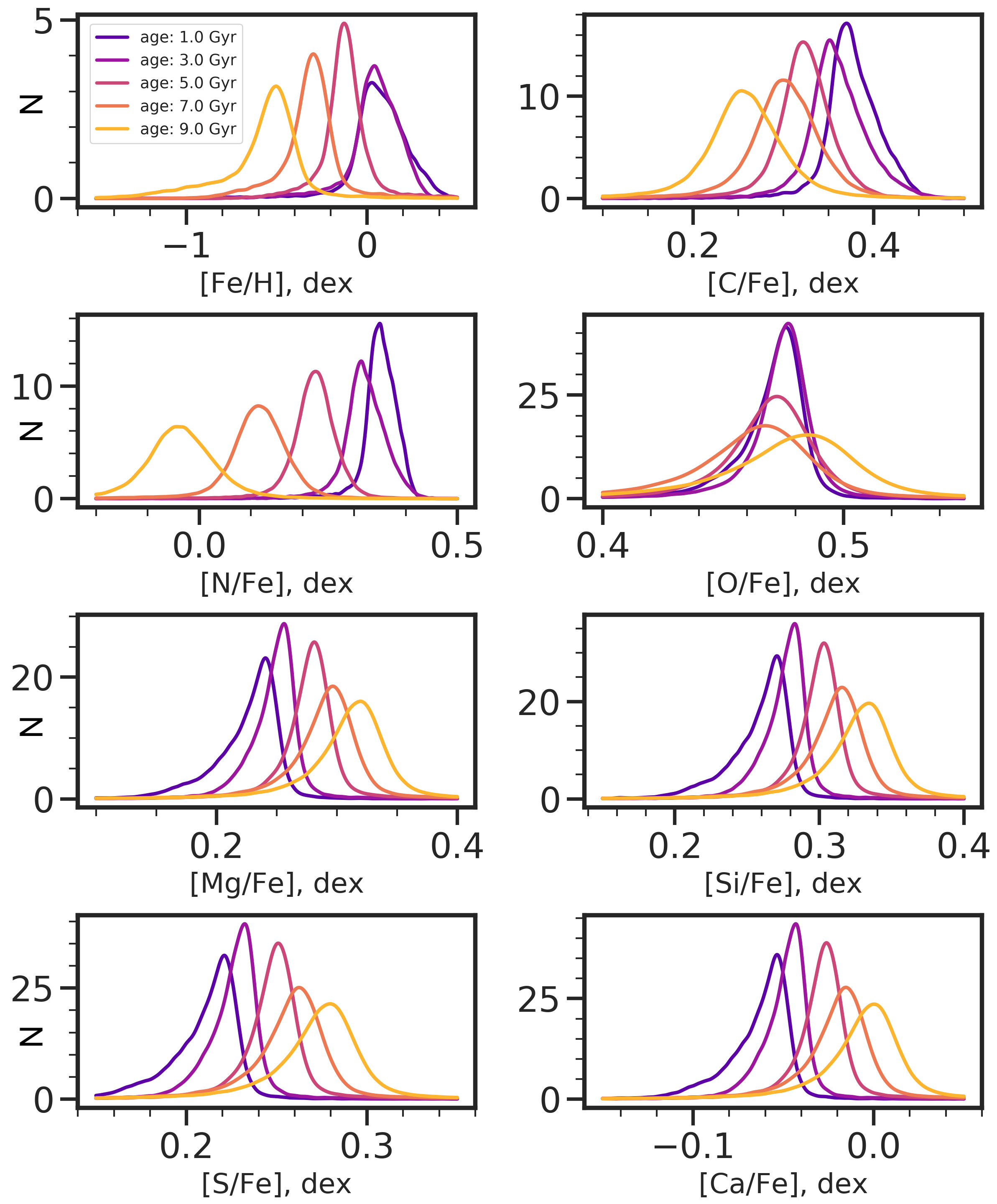} 
\caption{\textit{[X/Fe] distributions for stars at different ages.} KDEs for the [X/Fe] tracked in this study with legends similar to Figure \ref{fig:R_kdes} i.e., purple to yellow go from 1 Gyr to 9 Gyr stellar populations. The KDEs are similarly normalized to the number of stars per age group. All elements show distinct [X/Fe] distributions for different ages, except for O, where the different distributions show substantial overlap. } 
\label{fig:chem_R_kdes}
\end{figure*}

A starting point to understanding the present day distributions of abundances across the disk (see Section \ref{sec:abund_age_trends}) is to look at the birth radius distributions for different age population.
In Figure \ref{fig:R_kdes}, we show the kernel density estimation (KDE) for the birth radius, $\rm R_{birth}$,  
of stars centered on different age bins from ages =  1 to 9 Gyr. Because the number of stars within each age bin is different, we normalize the KDEs such that the distributions are on the same scale for easier comparison. We used the smallest possible bin in age such that the measured dispersion across the [Fe/H] vs  $\rm R_{birth}$ trend is not induced by the age binning, giving age bin sizes of (0.063, 0.125, 0.063, 0.125, 0.250) Gyr for the (1, 3, 5, 7, 9) Gyr samples, respectively. For more details on this process, we refer to Appendix \ref{sec:appendix_bin}. We applied this same age binning in comparing the KDEs at different ages for 
[Fe/H] and [X/Fe], where X={C, N, O, Mg, Si, S, Ca} shown in Figure \ref{fig:chem_R_kdes}.

The $\rm R_{birth}$ distributions peak from the outskirts of the galaxy to the central regions, from the youngest to the oldest stars, respectively. However, this is not observed for the 9 Gyr stellar population, which is more spread out across the disk. This is because at higher redshifts (i.e. lookback time $>$ 8 Gyr), the star formation in m12i was violent and bursty until a final minor merger finishes at lookback times of $\sim$6 Gyr, after which a more stable and calm disk appears and persists \citep{ma17}.  

Next, we focus on the  distributions of element abundances as a function of age as shown in Figure \ref{fig:chem_R_kdes}.
\\
\\
\noindent \textbf{Metallicity:} The [Fe/H] KDEs follow an increasing trend with decreasing age, though the 1 Gyr and 3 Gyr age bins have similar [Fe/H] distributions, and the width of the distribution also varies with different age bins. The similarity in the [Fe/H] distributions for the younger stars is in line with the little evolution in [Fe/H] and the similar radial abundance gradients at these ages found by \citet{bellardini22} in their sample of Milky Way-like galaxies in FIRE-2.  Qualitatively, the [Fe/H] distribution widths seem to be related to the $\rm R_{birth}$ distribution widths shown in Figure \ref{fig:R_kdes}, although we caution that this could be due to our age-binning, which was derived from the [Fe/H] vs $\rm R_{birth}$ trend at different ages. For the 9 Gyr bin, there is star formation everywhere in the galaxy, which induces a wide distribution in [Fe/H]. There is also a large tail towards lower metallicities reflecting the chemical enrichment at earlier times. Additionally, the large tail is affected by the merger with another smaller system, seen as a slight bump in the $\rm R_{birth}$ distribution at 20 kpc in Figure \ref{fig:R_kdes}. For the 7 Gyr and 5 Gyr age bins, their [Fe/H] distributions are narrower, following similarly narrow and more-peaked distributions in $\rm R_{birth}$. Lastly, for the 3 Gyr and 1 Gyr stellar populations, the [Fe/H] distributions are similar, peak above solar values, and have broader distributions, reflecting the star formation that is happening in a wider range of $\rm R_{birth}$ across the disk. In observations, there is a known age-metallicity relationship in Milky Way disk stars \citep{soubiran08}, that shows decreasing [Fe/H] with older ages. On the other hand, many studies \citep{bergemann14,rebassamanserga16} have also found that there is no clear trend in this relationship. In a sense, the [Fe/H] distributions in this work agree with both scenarios, with the important distinction being the age at which this holds true. In this work, we find that the 1 and 3 Gyr age bins show little change in their respective [Fe/H] distributions, while increasing in age from the 5 Gyr bin shows progressively more metal-poor distributions. \citet{bellardini22} shows this more clearly for their sample of Milky Way-like galaxies in FIRE-2 in looking at the age-metallicity relationship in the simulations (see their Figure 1). The [Fe/H] shows no trend with age for the youngest stars up to $\sim$4 Gyr, after which there is a slight decrease in [Fe/H] with increasing age up to 7 Gyr, and a more drastic decrease in [Fe/H] at much older ages.   
\\
\\
Looking at the trends in [X/Fe] distributions with age in Figure \ref{fig:chem_R_kdes}, we find some elements follow expectations from observations while others diverge. We discuss these [X/Fe] distribution trends in order of atomic mass, from the light elements to the $\alpha$-elements. We also list the summary statistics of abundance distributions in Table \ref{tab:kdes} and where available, we included summary statistics from observations of the Milky Way low-$\alpha$ disk taken from \galah~DR3 written in brackets. We note however that the age bins used for the observations are much bigger (e.g., $\pm$0.8 Gyr) to take into account age uncertainties. Due to a relatively smaller sample, we also did not include a 1 Gyr bin from the observations. 
\\
\\
\noindent \textbf{Carbon:} The carbon abundance, [C/Fe], increases in peak [C/Fe] value with younger populations in the simulation. In this work, we have shown that stars in different age bins have roughly different [Fe/H] distributions in the -1 $<$ [Fe/H] $<$ 0 dex range; with this relation, we find that the [C/Fe] distributions in the simulations are contrary to what is found in observations, wherein [C/Fe] decreases with increasing metallicity at [Fe/H]$>$-1 dex (e.g. \citealt{bensby06}). This is also supported by our comparison to the \galah~data as listed in Table \ref{tab:kdes}, where the mean [C/Fe] increases with older ages in the observations but decreases in the simulations. C is mostly produced in massive stars, followed by low-mass AGB stars \citep{kobayashi20} in the Galaxy. In the simulations, there is a larger contribution of C coming from stellar winds, which is one possible source of discrepancy. In observations, the atmospheric abundance of C is also affected by dredge-up in convective stars. Throughout the life of a star, its C+N+O abundances remain relatively constant. However, the slowest reaction in the CNO cycle produces a net increase in $\rm ^{14}N$, decrease in $\rm ^{12}C$, and slight decrease in $\rm ^{16}O$ \citep{Masseron2015, martig16}. This is especially seen with RGB stars that have deep convective zones which take materials formed in the core onto the surface of the star, changing the atmospheric abundance. However, in the simulations, we have yields (not necessarily atmospheric abundance) which are IMF-averaged and all stars generated from a star particle inherit the same chemistry. Therefore, although this process is prescribed on the population level in the simulations (i.e., from \citealt{vdhoek97}, \citealt{marigo01}, \citealt{izzard04}), the abundances for individual stars are more nuanced than can be achieved by this procedure.
\\
\\
\noindent \textbf{Nitrogen:} The nitrogen abundance, [N/Fe] distribution distinctly changes with the stellar population age, from broader with lower [N/Fe] abundances at older ages and narrower with higher [N/Fe] abundances at younger ages. The [N/Fe] distributions at different ages also overlap the least among the elements considered in this work, which is a result of its yield's metallicity-dependence from SNe II as implemented in the simulations \citep{hopkins18}. In FIRE-2, only the elements N and O have yields from progenitors with metallicity-dependence, which is linear with progenitor [Fe/H] up to $Z/Z_{\odot}$ = 1.65. N is also intricately linked to C through the CNO cycle. However, compared to C, N is mainly produced in intermediate-mass AGB stars \citep{kobayashi11}. Like C, the effects of dredge-up on the N abundance is applied in the simulation yields, although IMF-averaged. 
Works using large spectroscopic survey data like \apogee~(e.g. \citealt{hayes18}) find a decrease in [N/Fe] with increasing [Fe/H] up to [Fe/H]$\sim$-0.5 dex, after which [N/Fe] increases with increasing [Fe/H], similar to what we find in the simulations. 
\\
\\
\noindent \textbf{Oxygen: }The oxygen abundance, [O/Fe], in Ananke shows the smallest evolution across time among the elements in this study i.e., the [O/Fe] distributions show the most insignificant differences for stars of different ages. The [O/Fe] distributions are wide at older ages and become narrow at younger ages, but, with each distribution overlapping substantially. In the simulations, most of the O is produced via SNe II which is expected for this hydrostatic $\alpha$-element. However, in FIRE-2, there is also non-negligible contribution from stellar winds with a dependence on the progenitor metallicity. The stellar winds contribute a comparable amount of O to the ISM as SNe II at 3 $\times$ solar [Fe/H]. This trend in the simulations is contrary to what is found in observations, where [O/Fe] is seen to decrease with increasing [Fe/H] (e.g. \citealt{bensby04}), and loosely, age, as is expected of an $\alpha$-element.  This increase in abundance with older age is indeed what we also find for the [O/Fe] distribution trends in \galah. The significant O production from stellar winds at higher metallicities is a possible source of discrepancy.  
\\
\\
\noindent \textbf{$\alpha$-elements (Magnesium, Silicon, Sulfur, Calcium): }Lastly, we consider the other $\alpha$-element abundance distributions i.e. Mg, Si, S, and Ca. Although these elements have distributions with different mean values, the characteristics of the distributions for the different age bins are similar: broad at older ages with higher abundances, evolving to narrow peak at younger ages with lower abundances. This behavior we see for the mean [X/Fe] is expected for $\alpha$-elements, which are mainly produced through SNe II with shorter timescales, and diluted via SNIa that have delayed and longer timescales, therefore decreasing the $\alpha$-abundance with younger generations of stars. This is also what we observe in the \galah~data, aside from Ca, which is intriguingly constant with age in observations (see also discussion in Section \ref{sec:abund_age_trends}), but increasing with older ages in the simulations like the other $\alpha$-elements. Mg, like O, is a hydrostatic $\alpha$-element produced during the shell burning phase in massive stars. Si, S, and Ca are explosive $\alpha$-elements that also have non-negligible contributions from SNIa. For the explosive $\alpha$-elements, the peaks in the [X/Fe] distributions between the oldest and youngest stellar populations are separated by only $<$0.1 dex in the simulations. On the other hand, the same mono-age populations are more widely separated for Mg that has negligible contributions from SNIa. In the simulations, Mg and Si are produced in similar amounts from SNe II, with S yields being 1/3rd and Ca yields being 1/25th of the Mg yields. 
Among the $\alpha$-elements, SNIa contributions are highest for Si and S, and lower by a factor of 13 and 20 for Ca and Mg, respectively. We also note a tail in the distributions of these elements towards lower [X/Fe], at all age bins, and most pronounced at 1 Gyr. 

\begin{table*}
\begin{center}
\caption{\textit{Summary Statistics of [X/Fe] Distributions}.  The mean ($\mu$) and standard deviation ($\sigma$) of each abundance distribution have been computed with standard \texttt{numpy} functions, while the skew and kurtosis have been calculated with \texttt{scipy.stats}. The $\mu$ and $\sigma$ have units of dex while the skew and kurtosis are unitless. The skewness is a measure of symmetry while kurtosis, though qualitatively a measure of how peaked a distribution is, is sensitive to outliers. Both skew and kurtosis are therefore affected by the tails in the distribution. For the skew, a value of zero is perfectly symmetric and resembles a Gaussian, negative values indicate more weight on the left tail of the distribution, and larger positive values indicate more weight on the right tail. For kurtosis, larger values indicate larger tails in the distribution or a narrower distribution, which is what we see in the simulations. We refrain from making conclusions based on these skew and kurtosis values, and mainly point out that the large values for the kurtosis and deviation from 0 for the skew, indicate the heavy tails in the [Fe/H] and [X/Fe] distributions inherited from the range of stellar $\rm R_{birth}$ at a given age. Where available, we included the values from observations from \galah~DR3, enclosed in brackets. }
\begin{tabular}{cccccc}
\hline \hline
  & 1 Gyr & 3 Gyr & 5 Gyr & 7 Gyr & 9 Gyr \\
\hline
 [Fe/H], mean ($\mu$, dex) & 0.07 & 0.04 [-0.08] & -0.12 [-0.10] & -0.32 [-0.14] & -0.58 [-0.21]\\
 scatter ($\sigma$, dex) & 0.17 & 0.15 [0.19] & 0.14 [0.23] & 0.17 [0.26] & 0.24 [0.31]\\
 skewness ($S_{k}$) & -1.55 & -1.44 [-0.48] & 0.45 [-1.03] & 0.37 [-1.64] & -0.76 [-2.79]\\
 kurtosis ($\kappa$) & 6.62 & 4.76 [2.94] & 5.69 [5.43] & 5.05 [8.56] & 3.58 [13.46]\\
\hline
 [C/Fe], $\mu$, dex & 0.37 & 0.36 [0.02] & 0.32 [0.02] & 0.30 [0.04] & 0.25 [0.08] \\
  $\sigma$, dex &  0.03 & 0.04 [0.12] & 0.05 [0.12] & 0.06 [0.13] & 0.06 [0.14] \\
  $S_{k}$ & -2.12 &  -2.23 [0.66] & -3.46 [0.59] & -2.31 [0.62] & -2.02 [0.91]\\
  $\kappa$ & 18.4 & 25.08 [2.41] & 31.75 [3.22] &  27.50 [1.79] & 24.32 [2.43]\\
\hline
 [N/Fe], $\mu$, dex & 0.35 & 0.32 & 0.22 & 0.11 & -0.04 \\
  $\sigma$, dex & 0.04 & 0.05 & 0.05 & 0.07 & 0.09 \\
  $S_{k}$ & -3.73 & -2.25 & -2.84 & -1.82 & -1.55 \\
  $\kappa$ & 35.90 & 22.62 & 26.69 & 18.37 & 14.19\\
\hline
 [O/Fe], $\mu$, dex & 0.47 & 0.47 [0.05] & 0.47 [0.04] & 0.46 [0.09] & 0.47 [0.16] \\
  $\sigma$, dex & 0.03 & 0.03 [0.17] & 0.04 [0.16] & 0.05 [0.17] & 0.06 [0.19] \\
  $S_{k}$ & -7.33 & -7.55 [0.35] & -6.02 [0.25] &  -5.36 [0.62] & -5.56 [0.89]\\
  $\kappa$ & 96.98 & 137.38 [3.95] & 74.92 [6.68] & 62.39 [5.57] & 71.02 [5.25]\\
\hline
 [Mg/Fe], $\mu$, dex & 0.22 & 0.25 [-0.04] & 0.27 [0.00] & 0.29 [0.03] & 0.31 [0.05] \\
  $\sigma$, dex &  0.04 & 0.03 [0.08] & 0.04 [0.07] & 0.05 [0.08] & 0.05 [0.08] \\
  $S_{k}$ & -3.64 & -4.91 [-0.14] & -5.90 [0.68] & -5.48 [0.68] & -5.78 [0.82] \\
  $\kappa$ & 36.48 & 86.24 [4.21] & 73.79 [3.27] & 64.86 [5.73] & 75.11 [ 7.58] \\
\hline
 [Si/Fe], $\mu$, dex & 0.26 & 0.28 [0.01] & 0.30 [0.02] & 0.31 [0.04] & 0.32 [0.07]\\
  $\sigma$, dex & 0.03 & 0.02 [0.06] & 0.03 [0.06] & 0.04 [0.07] & 0.04 [0.09]\\
  $S_{k}$ & -2.88 & -4.06 [1.32] & -5.26 [0.70] & -4.44 [1.50] & -4.38 [2.08]\\
  $\kappa$ & 24.36 & 61.09 [36.74] & 60.45 [18.57] & 43.25 [16.77] & 42.38 [14.49]\\
\hline
 [S/Fe], $\mu$, dex & 0.21 & 0.23 & 0.25 & 0.25 & 0.27 \\
  $\sigma$, dex & 0.02 & 0.02 & 0.03 & 0.04 & 0.04 \\
  $S_{k}$ & -2.66 & -3.80 & -5.05 & -4.16 & -4.05 \\
  $\kappa$ & 21.55 & 55.67 & 56.57 & 38.49 & 36.47 \\
\hline
[Ca/Fe], $\mu$, dex & -0.06 & -0.05 [0.07] & -0.03 [0.06] & -0.02 [0.05] & -0.01 [0.06] \\
  $\sigma$, dex & 0.02 & 0.02 [0.09] & 0.03 [0.08] & 0.03 [0.08] & 0.03 [0.09]\\
  $S_{k}$ & -2.47 & -3.53 [-0.19] &  -4.81 [0.22] & -3.92 [-0.03] & -3.78 [-0.24]\\
  $\kappa$ & 19.20 & 50.59 [4.99] & 52.24 [4.22] & 34.74 [5.26] & 32.04 [19.88] \\
\hline
\hline
\end{tabular}
\label{tab:kdes}
\end{center}
\end{table*}

In this section, we found that mono-age populations in the thin disk of m12i within $\rm R < 25 kpc$ trace birth locations (Figure \ref{fig:R_kdes}) that overlap quite broadly but peak at different $\rm R_{birth}$ and exhibit [Fe/H] and [X/Fe] distributions that also show distinct trends (Figure \ref{fig:chem_R_kdes}). Interestingly, the [Fe/H] trends mirror $\rm R_{birth}$ trends, where a wider $\rm R_{birth}$ distribution also corresponds to a wider [Fe/H] distribution; together, age and [Fe/H] trace $\rm R_{birth}$ in the disk of m12i. 

Irrespective of the [Fe/H] distributions, however, the width of the [X/Fe] distributions increases at older stellar ages. The oldest stars (ages $>$9 Gyr) were born from less homogeneous gas, when star formation was happening more chaotically in the galaxy as seen in previous works (e.g. \citealt{ma17}, \citealt{agertz21}, \citealt{bellardini21}). This is further supported by the larger azimuthal abundance scatter at higher lookback times for similar Milky Way-like galaxies in FIRE-2 \citep{bellardini22}. This subsequently produces wider abundance distributions. The increasing kurtosis in [X/Fe] with younger stellar ages, best seen in the elements N and O, indicates that the chemical enrichment in the disk becomes more homogeneous over time. At the same time, the skew in the [X/Fe] distributions becomes more pronounced at younger ages, which could be explained by \textit{where} the stars were born. Focusing on the 1 Gyr population, part of the sample was born at smaller radii, from gas that has had further dilution from SNIa (i.e., greater overlap with older populations as seen in Figure \ref{fig:R_kdes}), therefore skewing the [X/Fe] distribution more negatively. However, these populations can be disentangled with [Fe/H], which would tell us which stars were born more centrally (i.e. higher [Fe/H]) or in the outskirts (i.e. lower [Fe/H]). Therefore, age is indeed intricately linked to the  [Fe/H], [X/Fe], and $\rm R_{birth}$ of a star.

In this section, we also showed that the trends in the [X/Fe] distributions with age for the elements N, Mg, Si, and S in the simulations agree with expectations from observations, while those for the elements of C, O, and Ca do not. 
These similarities and differences have repercussions in the resulting age-[X/Fe] trends that we discuss in the following section.

\section{Age-abundance-present-day-location trends at fixed [Fe/H]}
\label{sec:abund_age_trends}

\begin{figure}
\centering
\includegraphics[width=0.4\textwidth]{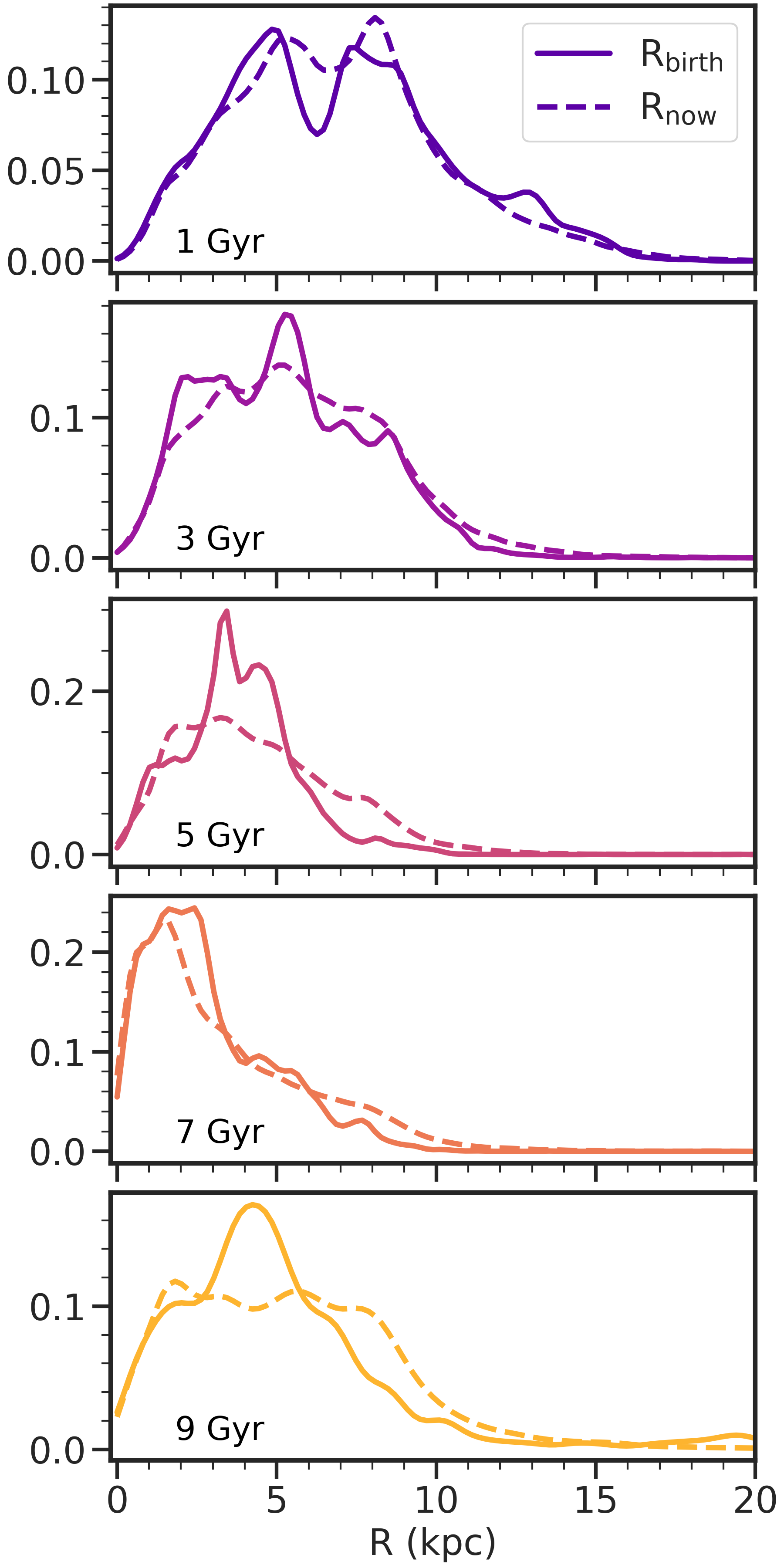} 
\caption{\textit{Current and birth radius for stars of different age ranges}. Kernel density estimations for $\rm R_{now}$ (solid) vs $\rm R_{birth}$ (dashed) 
of stars within age bins centered at 1 Gyr (purple) to 9 Gyr (yellow) in steps of 2 Gyr (top to bottom). The KDEs are normalized to the number of stars per age group. The location KDEs, both for $\rm R_{now}$ and $\rm R_{birth}$, exhibit similar tracks for stellar populations of the same age. This shows that although stars move from their birth location as shown by the changing R distributions, the bulk expectation for an inside-out galaxy growth is still observed, with old stars concentrated towards the center and young stars in the outskirts.  } 
\label{fig:Rnow_kdes}
\end{figure}

We have seen in the prior section the spatial trends of stellar element abundances, that are imprinted at birth. However, stars of different ages will have since moved from their birth location, so any initial trends will not be trivially maintained. As shown in Figure \ref{fig:Rnow_kdes}, there is a smoother distribution of current stellar location, $\rm R_{now}$ compared to $\rm R_{birth}$, as blurring and churning (i.e. radial migration) processes wash out the original clumpiness that stars were born into (e.g., \citealt{frankel20}). However, both $\rm R_{now}$ and $\rm R_{birth}$ 
still show distinct trends at different age groups with younger stars in the outskirts and older stars more centrally-located. 

In this Section, we examine what the current distribution of element abundances at different locations and [Fe/H] as a function of age looks like, as a result of the combination of birth properties and migration. We note that from within the cuts outlined in Section \ref{sec:sample_selection}, we pivot to two fixed [Fe/H] values, of [Fe/H] = 0 dex and [Fe/H] =  $-0.25$ dex, for the purpose of demonstrating the age--[X/Fe] trends for a high and a low [Fe/H] sample, respectively.

\subsection{Solar neighborhood at solar [Fe/H]}
\label{sec:solmet_solneighborhood}

\begin{figure*}
\centering
\includegraphics[width=\textwidth]{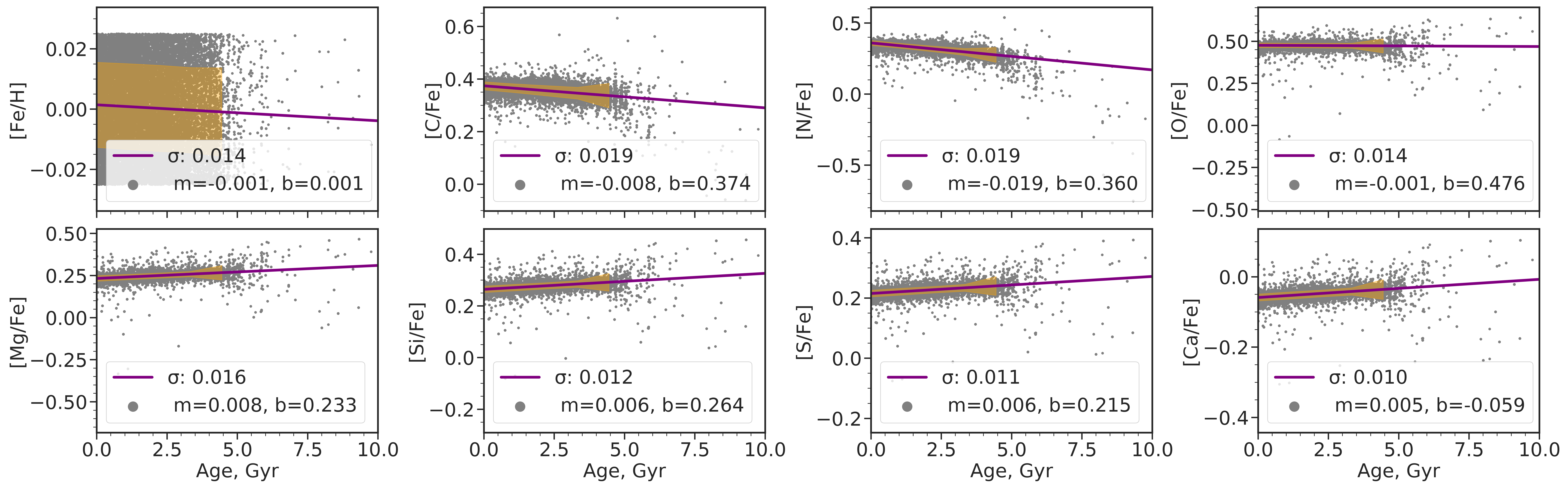}
\caption{\textit{Age-individual element abundance relationships at \textbf{[Fe/H] = 0 dex }in the solar neighborhood}. [Fe/H] and [X/Fe] (y-axis) as a function of age in Gyr (x-axis), for the abundances provided in the simulations. Included in the plot are stars within 1 kpc of LSR2 in m12i with [Fe/H] = 0 $\pm$ 0.025 dex. We plot the best fit line (solid purple) and the 1-$\sigma$ running dispersion from the line (orange shaded region), which we mark on the legend of each subplot. The $\alpha$ element abundances (Mg, Si, S, Ca) increase with age, and C, N, O, and Fe decrease with age in the simulation. There is a very small intrinsic scatter around these age-[X/Fe] relations, of \sigint $<$ 0.02 dex. } 
\label{fig:ageabundancesolarmet}
\end{figure*}

First, we investigate the age-[X/Fe] trends, at solar metallicity ([Fe/H] = 0$\pm$0.025 dex) in the solar neighborhood (1 kpc from LSR2), as shown in Figure \ref{fig:ageabundancesolarmet}. We fit a linear function to the age-[X/Fe] trends and the best fit is shown as the purple solid line with a running dispersion shown as the orange shaded region. The quantities from these fits are included in each figure sub-panel i.e., the median intrinsic dispersion (\sigint), the slope, m, and the intercept, b. Similar to the age binning derivation in Section \ref{sec:kdes}, we determined the best binning in [Fe/H] by progressively calculating the \sigint~at different bin values and choosing the bin that gives the smallest value so that its measurement is the least sensitive to the size of the [Fe/H] bin. We note that observational studies of the age-[X/Fe] relations typically have fixed and larger [Fe/H] bins (e.g., 0.05-0.10 dex) compared to this work, but we have added the step of determining the best binning to remove the dependence of the \sigint~in [X/Fe] on the scatter in [Fe/H]. Nonetheless, the discussion on the \sigint~from simulations and observations in Section \ref{sec:obscomparison} show that even with different [Fe/H] binning, their comparison is still viable. We excluded the running value of the intrinsic dispersion, \sigint,~calculated for age-[X/Fe] bins with fewer than 500 stars, to avoid being skewed to larger \sigint~due to sparsity in certain regions of the age-[X/Fe] parameter space. However, we note that sparsity still affects some of our analysis.

The cuts we applied (see Section \ref{sec:obsdata}) allow for a more direct comparison of the age-[X/Fe] trends between simulations and observations. However, there are some discrepancies in the scaling of the [X/Fe] between observations and simulations. Given this, we focus more on the similarities and differences in the [X/Fe] trends \textit{with} age. Absolute differences in the mean element abundances between data and simulation are not unusual and could be caused by many things in the simulations, like the different star formation history in m12i compared to the Milky Way \citep{sanderson20}.

At [Fe/H] = $0\pm0.025$ dex, the age-[Fe/H] relationship is essentially flat (slope is -0.001 dex/Gyr), with a slight decrease at older ages, which is to be expected as older stars generally have lower [Fe/H]. We measure a \sigint~of 0.014 dex for the age-[Fe/H]. Note that this is an arbitrary \sigint~because we fixed the [Fe/H] bin size at 0.05 dex; it is merely a reflection of the [Fe/H] range at which we consider the other age-element abundance trends at this [Fe/H]. 

The age-[C/Fe] relationship shows decreasing [C/Fe] as a function of age. This is opposite to what is found in observations (e.g. \citealt{nissen15,bedell18,ness19}). As discussed in Section \ref{sec:kdes}, C is produced mainly in SNe II in observations, and as a consequence, should have higher abundance for older stars than younger stars. In fact, \citet{nissen15} found that [C/Fe] has a larger slope with age than $\alpha$-elements, and proposed that a top-heavy IMF can possibly explain the strong trend in C abundance with age. There is a small intrinsic dispersion around the age-[C/Fe] trend, of \sigint=0.019 dex. 


The age-[N/Fe] trend shows a decreasing trend with older ages, similar to what is found in observations of RGB stars from \apogee~\citep{ness19}. The measured slopes from the Milky Way observations and the simulations agree, with m=-0.019 \dexgyr~from this study (and \sigint=0.019 dex) and m=-0.02 \dexgyr~from \citet{ness19}. Unfortunately, N is hard to measure for stars in the optical so we are unable to compare to more observational studies.  

The [O/Fe] trend with age is slightly decreasing, but essentially flat, having m=-0.001 \dexgyr~and a \sigint=0.014 dex around the best fit line. This is contrary to what is observed for the Milky Way \citep{nissen15,bedell18,ness19,sharma22}. O is a hydrostatic $\alpha$-element dispersed mostly through SNe II and is expected to decrease with younger stars. As noted in Section \ref{sec:kdes}, the metallicity-dependence of O yields from stellar winds in the simulations significantly contribute at higher [Fe/H], therefore driving the trend to be flat or even decreasing with older age. 

All the remaining $\alpha$-elements---Mg, Si, S, and Ca---show positive age-[X/Fe] trends, in line with expectations from Galactic chemical evolution work \citep{kobayashi20} and from observations \citep{nissen15,bedell18,ness19,sharma22} with the exception of Ca, where observations find a flat age-[Ca/Fe] trend. Ca is an explosive $\alpha$-element, which is expected to have a similar slope as compared to other explosive $\alpha$-elements like Si and S. We observe this in the simulations where the slopes for Si, S, and Ca are 0.006, 0.006, and 0.005 \dexgyr~respectively, as the SNIa contributions flatten the trend with age, while in comparison, the hydrostatic $\alpha$-element Mg has m=0.008 \dexgyr, a steeper trend because it is mostly produced in SNe II.  Ca, aside from showing a flat relationship with age in observations also shows a flat relationship in stellar velocity dispersion with increasing abundance, contrary to what is found for other $\alpha$-elements \citep{conroy14}. This is especially intriguing because there is a known age-velocity dispersion relation in observations wherein stellar populations with older ages have higher stellar velocity dispersion. One way to explain this curious trend of Ca with age is to include contributions from other sources, such as a subclass of supernovae called ``calcium-rich gap transients" that would flatten the trend at later times \citep{mulchaey14,nissen15}.

For all the elements except Fe, the running \sigint~gets larger at older ages where there are (1) fewer stars in general at solar metallicity and (2) the stars are born from less homogeneous gas.   

\begin{figure*}
\centering
\includegraphics[width=\textwidth]{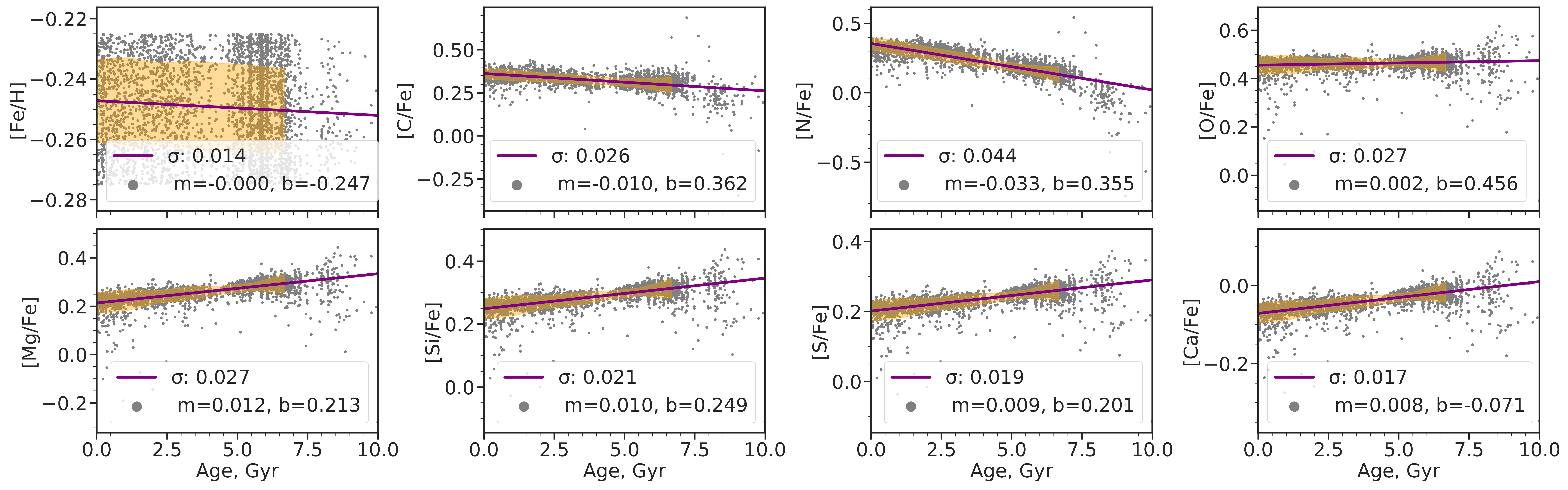}
\caption{\textit{Age-individual element abundance relationships at \textbf{lower metallicity, [Fe/H] = -0.25 dex}, in the solar neighborhood}. The legends and symbols are similar to Figure 
\ref{fig:ageabundancesolarmet}. All the elements at this lower [Fe/H] follow a similar trend to the solar [Fe/H] case with three main differences: (1) these age-[X/Fe] relations extend to older ages, (2) the trends are much steeper for all the elements (except for Fe), and (3) the scatter of the elements around their age-[X/Fe] relations, \sigint,~ are on average twice as large as at [Fe/H] = 0 dex (although still very small) with \sigint$<$ 0.045 dex.} 
\label{fig:ageabundancelowmet}
\end{figure*}

\subsection{Solar neighborhood at [Fe/H]= -0.25 dex}
\label{sec:lowmet_solneighborhood}
We apply the same spatial selection and binning determination as per Figure \ref{fig:ageabundancesolarmet}, but select stars at lower metallicity, with [Fe/H] = -0.25 $\pm$0.025 dex. We compare our age-[X/Fe] trends with \citet{ness19} and \citet{sharma22}, that also investigated at lower [Fe/H] i.e. $-0.35$ dex for the former and [Fe/H] $= -0.2$, $-0.4$, and $-0.5$ dex for the latter. We aim to compare not only the trends for simulations vs observations but also the age-[X/Fe] trend differences that are introduced at different [Fe/H]. As we noted in Section \ref{sec:sample_selection} and Figure \ref{fig:rnowhist}, for the simulated Milky Way-mass galaxy m12i, exploring at much lower metallicities than [Fe/H] $= -0.25$ dex probes stars that were born before the onset of a disk exhibiting distinct inside-out growth.   

\begin{figure*}
\centering
\includegraphics[width=1\textwidth]{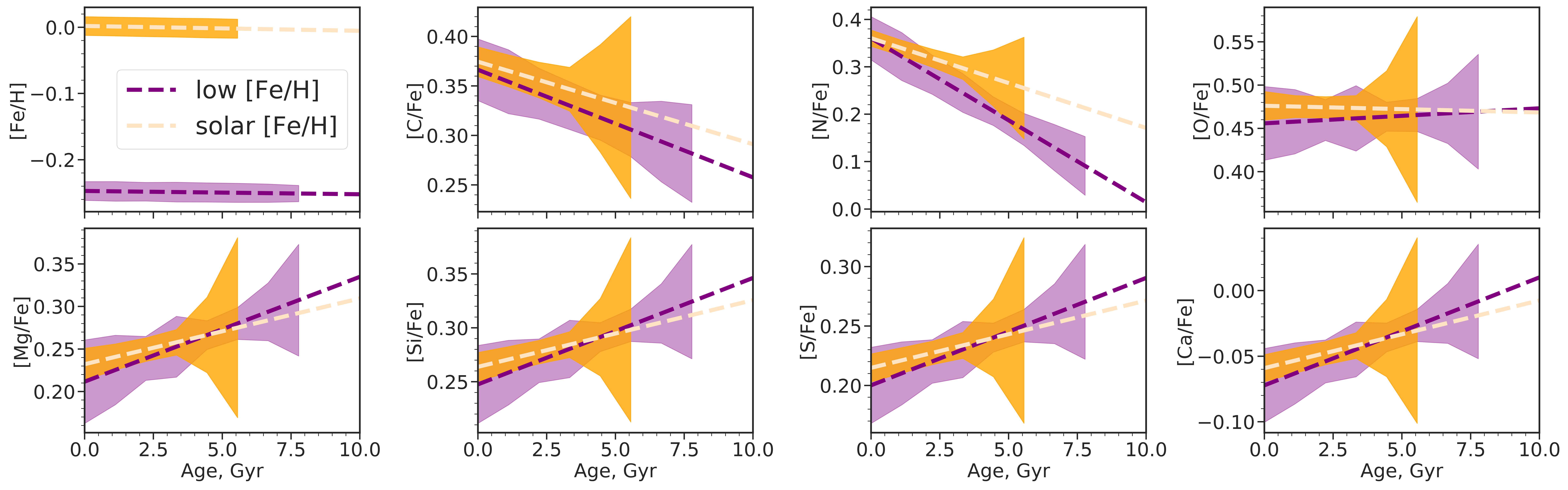}
\caption{\textit{Age-individual element abundance relationships at [Fe/H] = 0 dex and [Fe/H] = $-0.25$ dex in the solar neighborhood}. The running dispersions of the data around the age-[X/Fe] trends are taken from Figures \ref{fig:ageabundancesolarmet} and \ref{fig:ageabundancelowmet}. The age-[X/Fe] relations overlap and show flaring at older ages, especially for the solar [Fe/H] stars. However, this is mostly due to a smaller sample of stars at these ages.  } 
\label{fig:highlow_solarneighborhood}
\end{figure*}

Compared to the [Fe/H] = 0 dex sample, the stellar population at [Fe/H] = -0.25 dex extends to older ages, as is expected from the [Fe/H] KDE in Figure \ref{fig:chem_R_kdes}. Interestingly, the age-[X/Fe] trends at [Fe/H] = -0.25 dex in Figure \ref{fig:ageabundancelowmet} show more clumpiness and scatter than the [Fe/H] = 0 dex trend, especially at older ages i.e., one clump at 5$<$age$<$7.5 Gyr and another at age $>$8 Gyr. For each element, the age-[X/Fe] trend is steeper compared to the [Fe/H] = 0 dex sample (see Figure \ref{fig:highlow_solarneighborhood} for comparison), even without the inclusion of stars in the oldest clump. In fact, the age-[O/Fe] trend changes from slightly decreasing (or flat) at [Fe/H] $= 0$ dex to slightly increasing with a slope of m=0.002 \dexgyr, due to the less significant contribution from stellar winds at this point of m12i's chemical enrichment. The median \sigint~values are all higher for all the elements in this low-metallicity regime, except for Fe, where we made a cut in the sample selection. 

\citet{ness19} find that their solar and low [Fe/H] samples have comparable \sigint~and slopes. They find offsets in some age-[X/Fe] trends, especially for the $\alpha$-elements. In their observations, the trends for O, Mg, Si, S, and Ca at [Fe/H] = 0 dex are all lower compared to the trends at [Fe/H] = $-0.25$ dex. In the simulations, the trends for the solar and low [Fe/H] samples occupy a very similar space but they intersect at  $\sim$5 Gyr for the $\alpha$-elements, as shown in Figure \ref{fig:highlow_solarneighborhood}. However this could be easily explained by stars with different ages dominating the different [Fe/H] bins, where the median ages are 1 Gyr and 5 Gyr for the solar and low [Fe/H] samples, respectively. At the same age, the lower [Fe/H] sample is less chemically evolved and experienced less $\alpha$-element abundance dilution, making their age-[X/Fe] relations for the $\alpha$ elements higher compared to the solar sample. Additionally, a majority of the low [Fe/H] sample is older compared to the solar [Fe/H] sample, and these stars skew the age-[X/Fe] relations, resulting to the steepness in the trend (and therefore the cross-over). Additionally, for stars with ages $>$ 4 Gyr, the \sigint~in the solar [Fe/H] is higher than the \sigint~in the low [Fe/H] bin at a given age; however, we note that this is most likely due to the low number of stars at that age bin for the solar [Fe/H] sample compared to the low [Fe/H] sample. 
The observational work by \citet{sharma22} found that the age-[X/Fe] trends at different [Fe/H] overlap with each other substantially at all ages, similar to what we find. 

\begin{figure*}
\centering
\includegraphics[width=1\textwidth]{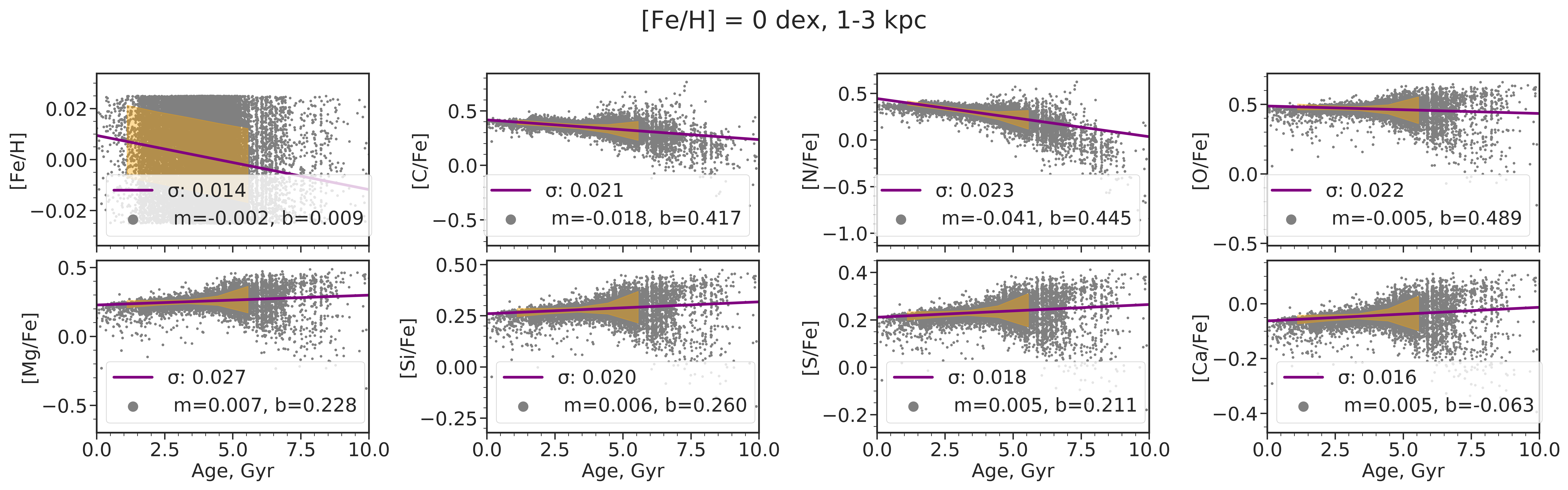}
\includegraphics[width=1\textwidth]{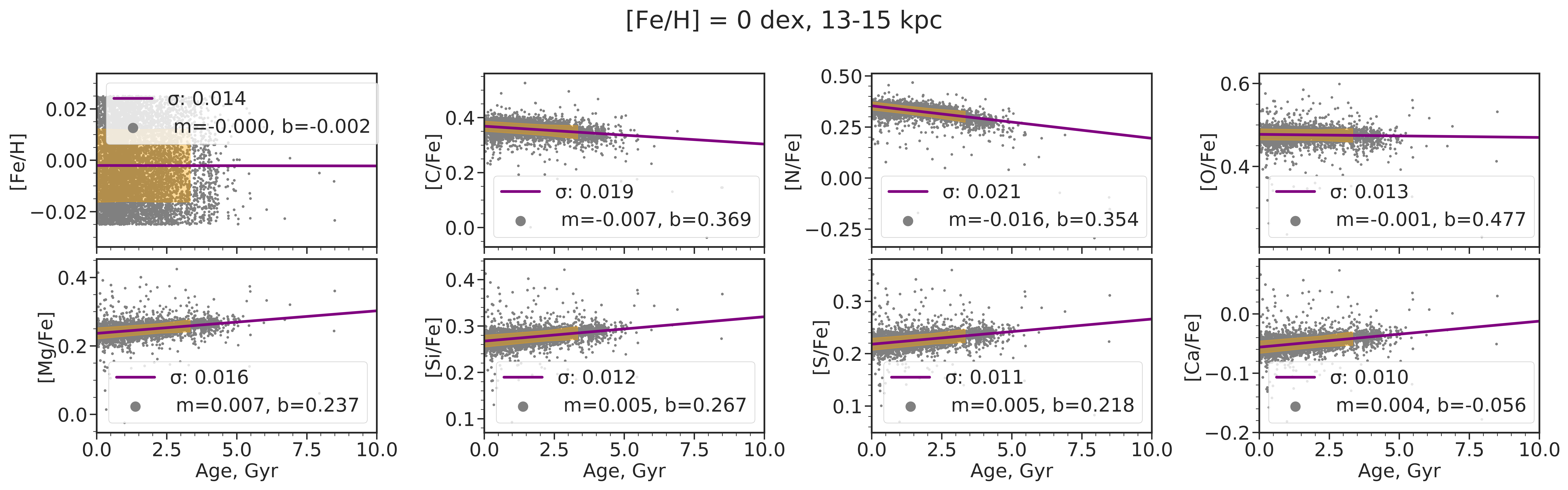}
\includegraphics[width=1\textwidth]{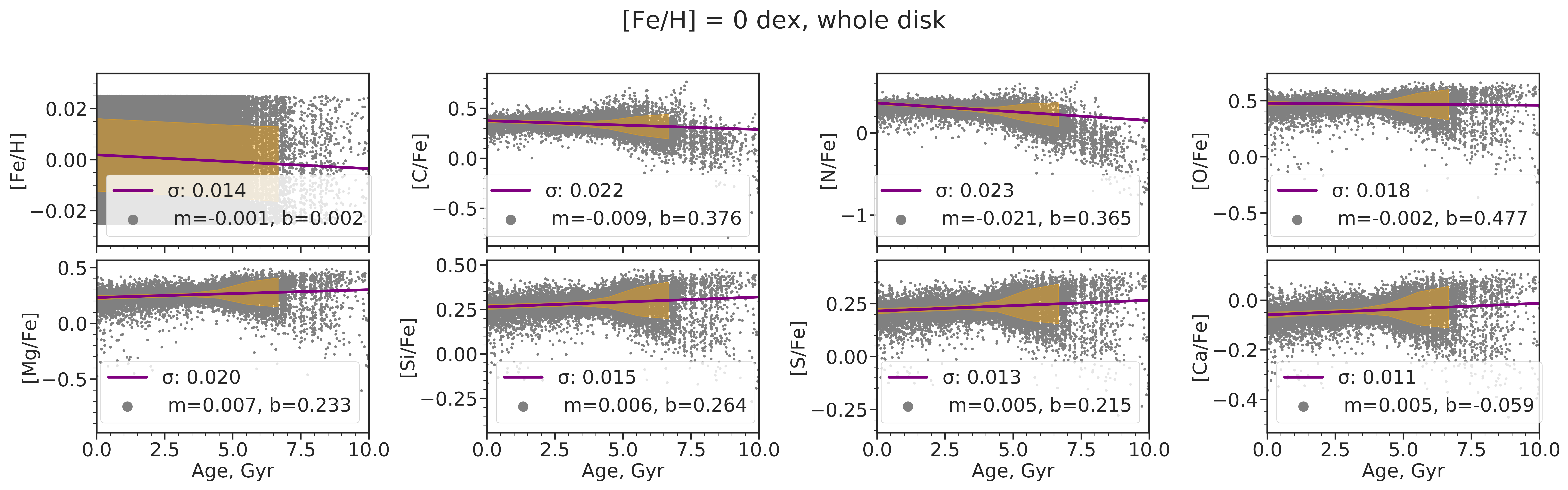}
\caption{\textit{Age-individual element abundance relationships at [Fe/H] = 0 dex in \textbf{different distance slices}}. All legends and symbols are similar to Figure 
\ref{fig:ageabundancesolarmet}. The top set of plots are stars in the inner disk at 1$\rm<R<$3 kpc, the center set are for stars in the outer disk at 11$\rm<R<$13 kpc, and the bottom set are for all stars within R = 25 kpc and Z$\pm$500 pc. The scatter around the age-[X/Fe] relations is higher for the inner disk compared to the outer disk.This is a consequence of older stars (of which there are more in the inner disk) exhibit wider abundance distributions. Interestingly, when all distance slices are included, the median \sigint~is still quite small at \sigint$<$0.025 dex.} 
\label{fig:ageabundancedistances}
\end{figure*}

\subsection{Different disk locations at solar metallicity, [Fe/H] = 0 dex}
\label{sec:solmet_diff_disklocations}

In this section, we explore the age-[X/Fe] trends in the simulations for stars at [Fe/H] = 0 $\pm$ 0.025 dex, Z $<$500 pc, and at different distance slices to represent the inner disk (1-3 kpc), outer disk (13-15 kpc), and the whole disk ($<$25 kpc) as shown in Figure \ref{fig:ageabundancedistances}. We particularly want to understand if different locations in the galaxy exhibit distinct age-[X/Fe] trends, and how this ties to the m12i's history. 

Compared to the solar neighborhood trend in Figure \ref{fig:ageabundancesolarmet}, the inner disk region (top set of subplots) shows higher \sigint~around the age-[X/Fe] relations. There is also a more pronounced increase in \sigint~for ages $>$ 4 Gyr. In fact, this behavior extends to much older ages where the number of stars is below the threshold that we have set. These old stars, as we have shown in Section \ref{sec:kdes}, and Figures \ref{fig:R_kdes} and \ref{fig:chem_R_kdes}, are born with a larger spread in their abundances when the star formation in the galaxy was more chaotic and clumpier. The stellar populations and the general age-[X/Fe] relations in the inner disk are both more complex, a reflection of the inner disk region's intricate history where stars of different ages all contribute. Nonetheless, we adopt a linear fit to the age-[X/Fe] distributions so as to enable more direct and interpretable comparison to observations and to the age-[X/Fe] trends at other locations. The increase that we see in \sigint, which signals the difference between steady and bursty star formation, is supported by the recent study by \citet{yu21}, which finds a similar transition at $\sim$3 Gyr for m12i.  This is only seen closer to the inner disk,  but is slightly noticeable in the solar neighborhood sample as well.

At the same [Fe/H] (i.e. [Fe/H] = 0 dex), the inner disk shows age-[X/Fe] relations in the $\alpha$-elements that are higher at older ages and lower at younger ages compared to the solar neighborhood (Figure \ref{fig:ageabundancesolarmet}) and outer disk (middle set of subplots in Figure \ref{fig:ageabundancedistances}). The higher $\alpha$-element abundance in the inner disk is in agreement with the rapid star formation and enrichment scenario in the center of galaxies, before the onset of SNIa.  On the other hand, the lower $\alpha$-element abundance in the same region is in line with the longer history of star formation in the inner regions of the galaxy where the [$\alpha$/Fe] abundances have been more diluted by SNIa compared to the outskirts of the galaxy. Compared to the age-[X/Fe] trends in the inner disk, the stars in the outer disk are mostly young with ages $<$5 Gyr and therefore have a more well-behaved age-[X/Fe] trend across the board that is well-fit by a linear trend.

Lastly, we look at the age-[X/Fe] trends for the entire disk (bottom set of subplots in Figure \ref{fig:ageabundancedistances}). Even with the inclusion of stars from various locations in the galaxy---from the inner disk that experienced different modes of star formation to the outer disk that has experienced later and more steady star formation---the median \sigint~remains small with values of \sigint $<$0.025 dex. We note that the gray data points are meant to show the total extent of the age-[X/Fe] trends, but not the density. This could be a misleading representation for the whole disk, where at ages $<$5 Gyr, the running \sigint~is small, but the extent in [X/Fe] at a given age is large. Nonetheless, this provides a direct comparison to the age-[X/Fe] relations discussed earlier. We see an increase in the running \sigint~towards older stars, bearing similarity to the inner disk. The small \sigint~measured in m12i, even after including all the stars ``observed" across the full disk sample points to a case where this small \sigint~is representative of a population whereby knowing the age and [Fe/H] of a star can predict its abundance, with increasing scatter at older ages. Ultimately, the ability to predict the abundance from the age and [Fe/H], highly depends on when the galaxy changed to a steady star formation from a more bursty mode as this determines the \sigint~and precision.

\section{Intrinsic Dispersion in the age-[X/Fe] -metallicity relation across the Galaxy's disk}
\label{sec:obscomparison}

\begin{figure*}
\centering
\includegraphics[width=\textwidth]{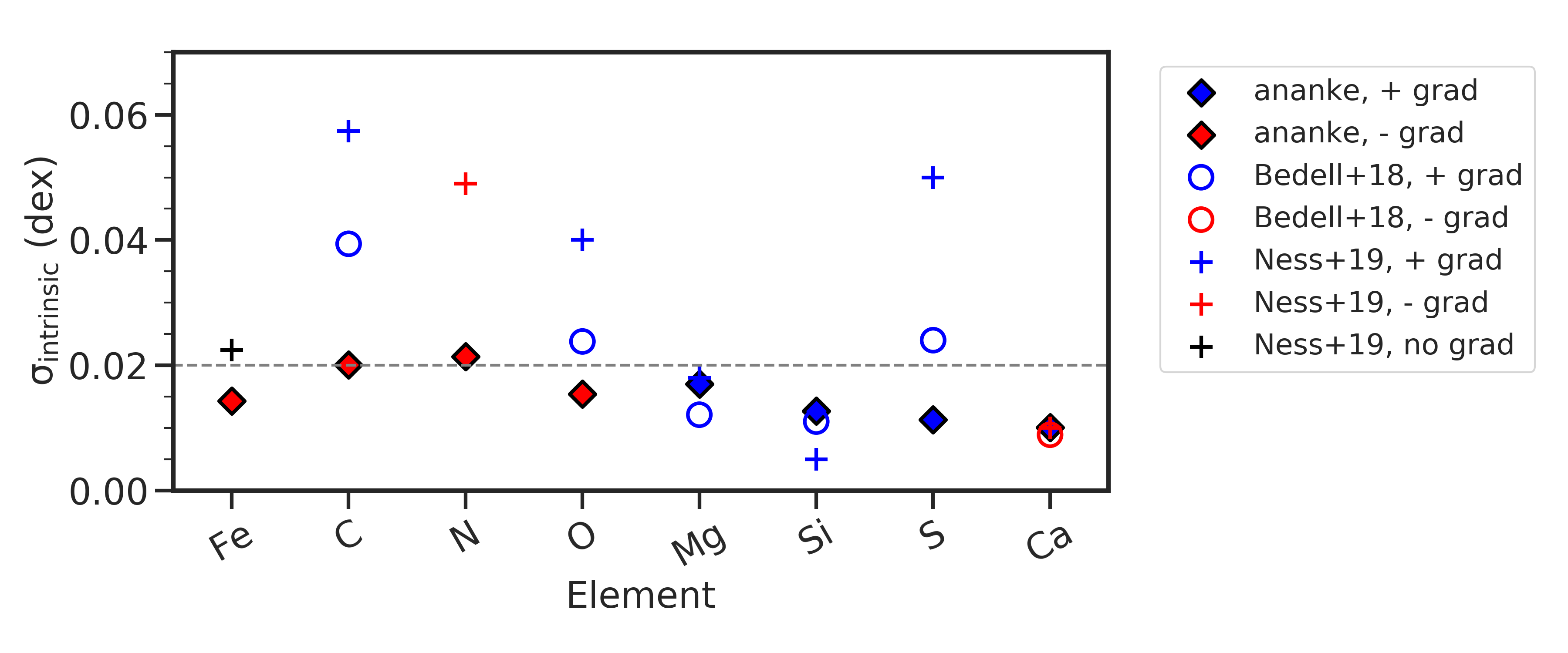} 
\caption{\textit{Intrinsic dispersions around the age-individual element abundance relations in simulations vs observations in the solar neighborhood}. We compare the \sigint~in the simulation (diamond) with those in observations: Solar twins from \citealt{bedell18} (circle) and red clump stars from \citealt{ness19} (cross), both at [Fe/H] = 0 dex. The symbols are also colored by the age-[X/Fe] slopes, with blue corresponding to increasing abundance with age and red to decreasing. $\rm \sigma_{int}$ = 0.02 dex is marked to guide the eye. In general, the $\rm \sigma_{int}$ from observations and simulations agree, barring C,N,O and S that have higher $\rm \sigma_{int}$ for the red giant stars. For the most part, the \sigint~tracked in the simulations is $<$0.02 dex. Aside from C,O, and Ca, the rest of the elements show similar trends in the simulations and observations.} 
\label{fig:dispersionsimvsobs}
\end{figure*}

We take a deeper look at the \sigint~in the simulation around the age-[X/Fe] relations, compared to Milky Way, as well as at different [Fe/H] and locations. We aim to see if this measurement of the scatter around the age-[X/Fe] relations can be used as a diagnostic and/or as a metric to capture the chemical enrichment in the simulated Milky Way galaxy m12i, and to 
relate this quantity to what is happening in the galaxy at that point in time. 

\subsection{\sigint~in simulation vs observations}

The intrinsic scatter around the age-[X/Fe] trends, \sigint, as measured for stars in the solar neighborhood at solar [Fe/H], from this work and other observational studies, is shown in Figure \ref{fig:dispersionsimvsobs}. The diamonds are stars from the simulations (Ananke), while the circles are solar twins \citep{bedell18}, and crosses are RGB stars \citep{ness19} from observations. As noted in Section \ref{sec:obsdata}, these observational data take into account the effects of the [X/Fe] and age uncertainties on the resulting \sigint. The symbols are colored by the age-[X/Fe] slope where red is negative and decreasing with age and blue is positive and increasing with age.  

The \sigint~for the elements produced in the simulated Milky Way are similar to the \sigint~from Milky Way observations, both for the solar twins and RGB stars, and are generally \sigint $\leq$0.02 dex, marked by the gray dashed line. This is especially true for the $\alpha$-elements Mg and Si. This small scatter is similarly found by \citet{bellardini22} for [Mg/Fe] at different age bins in their sample of FIRE-2 Milky Way-like galaxies. The general similarity in the \sigint~means that the level of homogeneity of star forming gas in the Milky Way at a given metallicity is well-reproduced in the simulated galaxy m12i. 

As discussed in Sections \ref{sec:kdes} and \ref{sec:abund_age_trends}, C, O, and Ca exhibit dissimilar trends with age compared to observations (and again note we refer to age-[X/Fe] relations in narrow bins of [Fe/H]). The element C shows a negative slope with age in the simulation, while it has a positive slope measured both from dwarf and giant stars in the Milky Way. The element Ca, on the other hand, has a positive slope, typical of $\alpha$-elements in the simulations. However, Ca has an observed negative slope with age. The element O is essentially flat with age in the simulations, in contrast to the increasing trend with age seen in observations. In a simple exploration of the age-[X/Fe] relations in the other galaxies in Ananke (m12m and m12f) with different star formation histories, we find that they similarly exhibit these discrepancies in the direction of the trends. This highlights the need for updates to the chemical enrichment sources, yields, and rates prescribed in the star formation in cosmological simulations in order to further their utility for Galactic archaeology purposes.  

There have been many recent efforts to look at the age-[X/Fe] trends in different Milky Way stellar populations or spectral types. These recent efforts are a reflection of how powerful a tool chemical abundances could be, especially for deriving ages (e.g. \citealt{hayden20}). \citet{lu21} compared the age-[X/Fe] trends for the high- and low-$\alpha$ disk using \apogee~red clump stars and found that the median \sigint~for both stellar populations are similar at \sigint $\sim$0.04 dex; even with stellar populations with different chemical enrichment histories, the abundance scatter is comparable. Although there is no $\alpha$-element dichotomy in m12i in the [$\alpha$/Fe]-[Fe/H] plane, we do find that the \sigint~measured at different radial parts of the disk are small and generally \sigint $<$0.04 dex, even if they experienced different formation histories. In this sense, our work agrees with the findings from \citet{lu21}. However, a further exploration of the age-[X/Fe] relations for different galaxy components, as \citet{nikakhtar21} recently determined for the Ananke galaxies, would be worth exploring in detail to compare to observations.  
Using \galah~DR3 data, \citet{sharma22} found that for ages $<$10 Gyr, a disk star's abundance can be predicted from its age and [Fe/H] with a \sigint~of 0.03 dex. They also found that comparing the slopes in the age-[X/Fe] trends to those from the [Fe/H]-[X/Fe] relations split elements into three groups, corresponding to their main channel of production: SNe II, SNIa, and stellar winds; these are the same sources used for the chemical enrichment in our simulated Milky Way. 

\begin{figure*}
\centering
\includegraphics[width=1\textwidth]{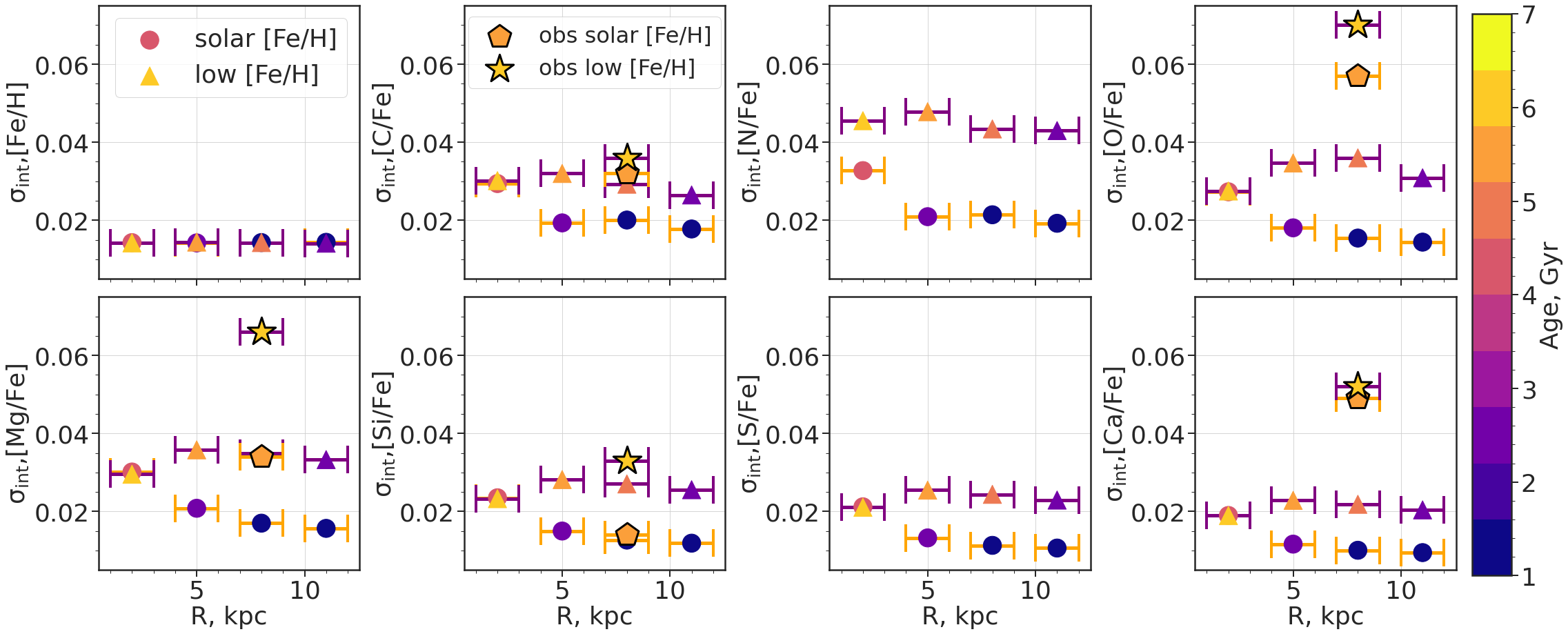} 
\caption{\textit{Intrinsic dispersion of individual element abundances [X/Fe] around the age-[X/Fe] trends at different disk locations and metallicities, for simulations and observations.} The \sigint~values are reported across different bins in galactocentric radii (1-3 kpc, 4-6 kpc, 7-9 kpc, and 10-12 kpc) for the elements in this study (solid symbols) as well as comparison with observations from GALAH DR3 (marked as a pentagon at solar [Fe/H] and star at low [Fe/H] in the 7-9 kpc bin for C, O, Mg, Si, and Ca). At each radius bin, we show the \sigint~for a solar metallicity sample (circles) and low-metallicity sample ([Fe/H] = -0.25 dex, triangles). The symbols are colored by the median stellar population age. Across all the elements for \textit{both} simulations and observations, the stars with lower [Fe/H] show higher \sigint~as well as higher median ages. There is also a trend of decreasing \sigint~with radius for the solar [Fe/H] sample which is not present in the low [Fe/H] sample.  } 
\label{fig:dispersioncomparison}
\end{figure*}

\subsection{\sigint~at different disk locations and [Fe/H] in the simulations}
\label{sec:sigint_simsvsobs}

We now explore the \sigint~of each element around their age-[X/Fe] relations for the seven elements included in this work (excluding [Fe/H]) at \textit{different radial locations} in the disk (e.g. 1-3 kpc, 4-6 kpc, 7-9 kpc, and 10-12 kpc) \textit{and at different metallicities} (solar at [Fe/H] $=$ 0 dex and low at [Fe/H] $=$ -0.25 dex) as shown in Figure \ref{fig:dispersioncomparison}. We color our markers by the median age of that stellar population in a given location and [Fe/H] bin. The 2kpc radial bin is consistent with the radial range over which the stars are similar in chemistry \citep{bellardini21}.

The \sigint~around the age-[X/Fe] trends differ for the two metallicity bins we examine, and this difference is larger than what is seen for the samples at different median ages or varying locations for stars at fixed [Fe/H]. 
The low-[Fe/H] sample, across all elements and radial bins, aside from the innermost region, exhibits higher \sigint~than the solar [Fe/H] sample. This is especially pronounced for the element N, which has metallicity-dependent yields from SNe II, thereby causing N to have the largest difference in \sigint~between the high and low [Fe/H] samples. In fact, only N shows a noticeable difference in the \sigint~at different [Fe/H] in the innermost radial bin. The element C shows the smallest \sigint~difference between the high and low [Fe/H] stars, meaning that a change in [Fe/H] produces minimal change in \sigint. The $\alpha$-elements i.e. O, Mg, Si, S, and Ca all generally show similar \sigint~trends with [Fe/H]: The solar [Fe/H] sample shows an exponential decline in \sigint~as a function of radial bin, while the low [Fe/H] sample does not exhibit a clear trend, and is mostly flat. 

As expected from how chemical enrichment proceeds and from Section \ref{sec:kdes}, the solar [Fe/H] stars have a younger median age compared to the low [Fe/H] stars at all radial bins. The median ages at a given [Fe/H] also vary depending on the location. For the solar [Fe/H] sample, the median age in the central region is $\sim$4 Gyr old, but for the outer radial bins, the median ages are $<$2 Gyr old. There is a larger range in median ages for the lower [Fe/H] bin, from 6.5 Gyr old in the central region of the disk, to 2.5 Gyr old in the disk outskirts. At the same radial bin, a larger difference in the median age between the two [Fe/H] bins does not correspond to a larger difference in \sigint.

In summary, we show that in comparing the variables of galactocentric radius, median age, and [Fe/H], the scatter of the elements, \sigint, around their age-[X/Fe] relations, changes the most with [Fe/H].  All the elements show differences in \sigint~between the low and high [Fe/H] bins, with a mean difference of $\sim$0.010 dex or a 60\% increase in \sigint~at low [Fe/H]. On the other hand, only for the solar [Fe/H] sample does the \sigint~vary with location, and at the same radius bin, the difference in median ages is not proportional to the difference in \sigint (for example, comparing between the 5 kpc and 11 kpc bin). Ultimately, there is the caveat that the \sigint~trends as a function of [Fe/H] and location in this work are unique to m12i. 

\subsection{\sigint~in the real solar neighborhood at different [Fe/H]}
\label{sec:}
In addition to the results from the simulation data, we derive the scatter around age-[X/Fe] relations, \sigint,~from \galah~DR3 observations of real solar neighborhood stars at different metallicities, shown in Figure \ref{fig:dispersioncomparison}. 
The following cuts were made to ensure that the stars from \galah~data (described in Section \ref{sec:obsdata}) are comparable to our bins of simulated stars and that the abundances are of good quality:

\begin{itemize}
    \item{7$<$R$<$9 kpc,}
    \item{signal-to-noise ratio $>$40,}
    \item{[Fe/H]=0$\pm$0.05 dex for the solar [Fe/H] bin,}
    \item{[Fe/H]=-0.25$\pm$0.05 dex for the low [Fe/H] bin,}
    \item{flag\_sp=0,}
    \item{flag\_X\_fe=0,}
    \item{5000$<$\teff$<$6100 K, and}
    \item{3.2$<$\logg$<$4.1 dex,}
\end{itemize} 
where X = C, O, Mg, Si, and Ca---the elements this study has in common with \galah. Note that our \teff~and \logg~cuts are adopted from the MSTO stars cut from \citet{sharma22}. With these selections, we end up with 2,882 stars for the solar [Fe/H] sample and 1,546 stars for the low [Fe/H] sample. In general, we performed a similar exercise as was done for the simulations, and calculated the median running dispersion in the best fit linear age-[X/Fe] trend for these elements. 

Like the other comparison observational data (see Section \ref{sec:obsdata}), the uncertainties in [X/Fe] and age have to be taken into account to calculate the \sigint. To account for the abundance uncertainties, we subtracted the mean $\rm \sigma_{abund}^2$ from the measured $\rm \sigma_{total}^2$ to get the $\rm \sigma_{int}^2$. The typical value for $\rm \sigma_{abund}$ is 0.05 dex for Mg, Si, and Ca and 0.10 dex for C and O. Next, to understand how the age uncertainty affects the measured \sigint, we drew a new age from a normal distribution centered on the age given by BSTEP with the age uncertainty as standard deviation. The mean age error for our sample is 0.7 Gyr. We do this 100 times and find that the change in  \sigint~is $\ll$0.01 dex and therefore negligible.  

We show the \sigint~in \galah~for stars in the solar neighborhood at  [Fe/H] = 0 dex and [Fe/H] = $-0.25$ dex, together with the \sigint~derived in the simulations in Figure \ref{fig:dispersioncomparison}. The [Fe/H] = 0 dex sample from \galah~has a median age of 5 Gyr, while the [Fe/H] = -0.25 dex sample has a median age of 6 Gyr. The \sigint~for the low [Fe/H] bin is higher than the solar [Fe/H] bin for all the elements with a mean offset of \sigint~0.014 dex, similar to what we find in the simulations. All measured \sigint~from the observations are lower than 0.07 dex, and the difference in \sigint~between the two [Fe/H] bins ranges from $<$0.01 dex (e.g. C, Ca) to $\sim$0.03 dex (e.g. Mg). The \sigint~trend for Si in observations resembles the \sigint~in the simulations the most,  while the \sigint~for Ca in \galah~resembles the simulations the least, where the \sigint~from the observations are larger by $\sim$0.03 dex compared to their simulated counterpart. Even with these differences, it is noteworthy that the \sigint~from the observations and simulations are very similar and of the same magnitude. This comparison highlights that even if the absolute values and trends for [X/Fe] in the simulations are dependent on many assumptions (i.e., on yields, rates, sources) and the star formation history, the general abundance scatter in observations is reproduced well \citep{hopkins18,escala18,bellardini21,bellardini22}. We expect that this agreement holds true even if realistic observational uncertainties were modeled for m12i as (1) the effect of the abundance error is removed when deconvolving the sources of the scatter and (2) the slopes in the age-[X/Fe] relations in the simulations are shallower than what the observations measure. Applying age uncertainties in the simulations will therefore have negligible effects on the resulting \sigint.  

\section{Discussion and Summary}
\label{sec:discussion_cca}

The stars in the Milky Way have been shown to exhibit tight age-[X/Fe] relations at a given [Fe/H], such that by knowing a star's age and [Fe/H], one can predict its abundance to a precision of $\sim$0.02 dex. Indeed, this has been studied in greater detail and in many axes e.g., high vs low-$\alpha$ disk,  high vs low [Fe/H], MSTO vs RGB stars, thanks to large spectroscopic survey data  \citep{ness19,lu21,sharma22}. 

The main goal of this work is to explore this observed age-[X/Fe] relations in cosmological zoom-in simulations that have been shown to reproduce Milky Way results from Gaia DR2 \citep{sanderson20}. One advantage to doing this is that we have absolute knowledge of when and where stars formed in the simulations. In contrast, ages in observations are harder to derive, the approach for deriving them is dependent on the type of star, and birth locations are unknown. Our main focus has been to determine if the current state-of-the-art simulations that trace the chemical enrichment of galaxies, can successfully reproduce the observed age-[X/Fe] trends in the Milky Way, and where they fall short and disagree. However, we can go beyond this and even explore what is physically happening in the simulated galaxy, and investigate where stars of a given age or chemistry are born.

Therefore, in understanding where the age-[X/Fe] trends in the simulations come from, we have also come to establish the intricate relationship of age, metallicity, individual element abundances, and location of stars in this simulated Milky Way galaxy, m12i. We have shown that aside from the chemical enrichment in the galaxy being a product of the prescribed nucleosynthesis (see Section \ref{sec:kdes}), the individual element abundances, and specifically their scatter around the age-[X/Fe] relations, \sigint, generally reflect m12i's star formation history. Focusing on the solar [Fe/H] sample, at all distances, we find that the running \sigint~across age is constant at young ages, up to about $\sim$4 Gyr, and then increases to higher \sigint~values at older ages. The detailed study of m12i by \citet{ma17} find that the disk changes from a bursty and clumpy star formation mode to a stable star formation mode at look back times of 6 Gyr. This galaxy was also included in the more recent analysis by \citet{yu21} where they found that m12i changes to a near constant state of star formation at $\sim$3 Gyr. Indeed, for stars that are $<$3 Gyr, the \sigint~is constant, for 3$<$age$<$6 Gyr, the \sigint~increases, and at $>$6 Gyr, the \sigint~is the largest. We would like to point out however, that we mainly see this behavior for the solar [Fe/H] sample. 

To make sense of the observed age-[X/Fe] trends, we have explored their distributions at different ages in Section \ref{sec:kdes}. Importantly, we also explored where these stars are born and where they are currently as shown in Figures \ref{fig:R_kdes} and \ref{fig:Rnow_kdes}. Here we find that stars in the oldest age bin (9 Gyr) are spread all across the disk within R $<$ 10 kpc, extending farther than stars that are younger in the 5 Gyr and 7 Gyr age bins. This might seem counter-intuitive in the inside-out formation scenario, where star formation starts in the central regions and proceeds outwards. However, for these oldest stars, the galaxy that eventually forms into m12i is composed of multiple smaller galaxies (see Figure 1 in \citealt{ma17}), which gives rise to the more extended and spread out $\rm R_{birth}$ distribution that we see. Indeed, \citet{santistevan20} found that the Milky Way/M31-mass galaxies in the FIRE simulations are made up of a hundred distinct dwarf galaxies before z$\sim$2. Therefore, the stars at the oldest age bin show the largest spread in the abundance distributions (as similarly seen in the analysis from \citealt{bellardini22} of the stellar abundances of Milky Way/M31-mass galaxies), which reflects the chemistry of this clumpier and more chaotic star formation, and with the [Fe/H] KDE showing a long tail towards lower values.  Curiously, the stars in the 3 Gyr bin have a similar $\rm R_{birth}$ (as well as current R) distribution to the 9 Gyr age bin, but its abundance distributions are narrower. This could be explained however by the existence of a rotationally-supported and stable disk that at 3 Gyr, has stars with abundances that are more azimuthally homogeneous at a given radius \citep{bellardini22}.

We also note that stars with different ages have significant overlap in their distributions of [X/Fe], $\rm R_{birth}$, and current R, but that the peak of those distributions are distinct from each other. For younger stars, the peak of the radial distributions of stars is at larger R, and the abundance distributions exhibit progressive enrichment. There is an overlap in the [X/Fe] KDEs, which is manifested in the gentle slopes in the age-[X/Fe] trends. The element N, whose yields from SNe II have a progenitor metallicity-dependence, has the most distinct [N/Fe] distributions as a function of age (i.e., [N/Fe] distributions overlap the least at varying age bins), and therefore also the steepest age-[X/Fe] trend. On the other hand, O, whose yields from stellar winds have a progenitor metallicity-dependence, has abundance KDEs that overlap for different ages, and therefore exhibits the flattest trend with age. We therefore find that \textbf{for the simulated Milky Way-like galaxy m12i, the \textit{direction} of the age-[X/Fe] trends is a reflection of the chemical evolution prescriptions in the simulations, and the \textit{scatter} around these relations is a reflection of the mode of star formation}, with higher scatter at earlier times when star formation was more bursty and chaotic, and lower scatter at later times when the disk is more stable. Meanwhile, the magnitude of the slopes in the age-[X/Fe] relations is a convolution of the effects from the chemical evolution prescription as well as the star formation history. 
With the next generation of spectroscopic surveys dedicated to densely mapping stars in the Milky Way such as SDSS-V \citep{kollmeier17},  WEAVE \citep{Bonifacio2016}, and 4MOST \citep{deJong2019}, as well as novel ways to derive ages for these stars, we can use the \sigint~in the age-[X/Fe] trend as another diagnostic in understanding the evolution of the Milky Way.

In Section \ref{sec:obscomparison} we compared the slopes of and \sigint~around the age-[X/Fe] relations from this work to those derived in observations. Specifically, in Figure \ref{fig:dispersionsimvsobs} we show the \sigint~in the solar neighborhood at solar metallicity for the stars in Ananke, solar twins \citep{bedell18}, and RGB stars \citep{ness19}. We find that the \sigint~for m12i and the Milky Way are similar, and in fact, some elements such as Mg and Si have near-identical \sigint~across all three samples. The elements C, O, and Ca, however, show the opposite trends in their slopes with age in the simulations compared to observations. The age-[C/Fe] relation increases with older ages in observations and has been noted to even have a steeper slope with age than $\alpha$ elements Mg and Si \citep{nissen15}. In contrast, we find that in m12i, the age-[C/Fe] relations have a negative slope;  decreasing with increasing age. We expect that this is because of the substantial contribution from stellar winds at later times prescribed in the simulations. 
The age-[O/Fe] relations show a positive slope in observations, as is expected for an $\alpha$-element. However, we find that the O abundance is flat, or decreasing with older ages in the simulations. This is due to the  large O yields from stellar winds at higher metallicities as prescribed in the simulations. Lastly, observations find that the [Ca/Fe] trend with age is flat or decreasing with older ages, contrary to the trend for the other $\alpha$ elements \citep{nissen15,bedell18,ness19,sharma22} and to what we find in this study, though this discrepancy could potentially be alleviated by the inclusion of other astrophysical sources (e.g., \citealt{mulchaey14}). The chemical enrichment in this cosmological zoom-in simulation of a Milky Way-like galaxy and observational comparisons highlight that our knowledge of how elements form and evolve through time is incomplete. Continued efforts for Galactic chemical evolution modeling (e.g. \citealt{kobayashi20}) in tandem with a more flexible way of propagating chemical enrichment are therefore important for these simulated galaxies to be an even more realistic laboratory for Galactic archaeology. Nonetheless, it is truly noteworthy that m12i, which was evolved in a large cosmological volume (later on zoomed-in which the Ananke synthetic survey was made from), and was matched as a Milky Way galaxy based \textit{only }on its mass and isolation, reproduce comparable age-[X/Fe] trends to the Milky Way. 

Moving forward, there are many avenues to delve deeper into Galactic archaeology in simulations, especially in the Ananke catalog, to link observed Milky Way properties to a plausible galaxy formation history. For example, it is worth looking more into the \sigint~for the different Milky Way galaxies in Ananke that have masses 1-2 $\times 10^{12} \rm M_{\odot}$ but have varying morphologies and star formation histories. This would be important in contextualizing what brings about the observed \sigint~in the Milky Way by comparing to the derived \sigint~in simulated Milky Way galaxies with different properties. In line with this, we have shown in Section \ref{sec:obscomparison} how the \sigint~for observations and simulations compare in the solar neighborhood, with the former calculated from \galah~abundances of Milky Way stars and the latter based on this analysis. With future surveys that will observe stars in the Milky Way more extensively (e.g. \citealt{kollmeier17}), we will be able to map the \sigint~in the Galaxy in its different regions, that have, in turn, their distinct stellar populations to aid in a more holistic picture of its formation history. \citet{bellardini21} used simulated Milky Way galaxies to show that the abundance scatter in gas is dominated by radial trends at lookback times $<$6.9 Gyr, and follow-up work show that this transition lookback time is $<$7.5 Gyr for stars \citep{bellardini22}, beyond which the abundance scatter is dominated by azimuthal inhomogeneity. It is therefore worthwhile to also explore the age-[X/Fe] trends from different solar viewpoints in the same galaxy to ultimately understand if the age-[X/Fe] relationship at a given metallicity is dominated by radial or azimuthal trends. \\
The future of the field is bright with large spectroscopic surveys in tandem with novel ways of deriving ages, as well as updates on hydrodynamical simulations to produce more realistic star formation in galaxies. This will allow the community to tackle how our Galaxy formed more expansively and from different perspectives, bridging our knowledge and the gap between observations and simulations, and with stars, specifically their chemical fingerprint, as the main tool. In this work, we give but a glimpse of this future, outlining how the abundance \sigint~can be a diagnostic for unraveling the evolution of the Milky Way. In the end, we are still left with many questions to investigate (e.g. galactic chemical evolution modeling, azimuthal vs radial abundance trends, flexible chemical prescriptions) both from the observational and simulation point of view, and we look forward to answering them in the future, in this era of big data in Galactic archaeology. 

\section{acknowledgments}
AC is grateful to the Flatiron Institute Center for Computational Astrophysics Pre-Doctoral Program through which this project was made possible. AC also acknowledges support from the Science and Technology Facilities Council (STFC) [grant number ST/T000244/1] and thanks the Large Synoptic Survey Telescope Corporation (LSSTC) Data Science Fellowship Program, which is funded by LSSTC, NSF Cybertraining Grant 1829740, the Brinson Foundation, and the Moore Foundation. MKN is in part supported by a Sloan Research Fellowship. AW received support from: NSF via CAREER award AST-2045928 and grant AST-2107772; NASA ATP grant 80NSSC20K0513; HST grants AR-15809, GO-15902, GO-16273 from STScI. KH and AC acknowledge support from the National Science Foundation grant AST-1907417 and AST-2108736 and from the Wootton Center for Astrophysical Plasma Properties funded under the United States Department of Energy collaborative agreement DE-NA0003843. This work was performed in part at Aspen Center for Physics, which is supported by National Science Foundation grant PHY-1607611. This work was performed in part at the Simons Foundation Flatiron Institute's Center for Computational Astrophysics during KH's tenure as an IDEA Fellow. RES acknowledges support from the Research Corporation through the Scialog Fellows program on Time Domain Astronomy, from NSF grant AST-2007232, from NASA grant 19-ATP19-0068, and from HST-AR-15809 from the Space Telescope Science Institute (STScI), which is operated by AURA, Inc., under NASA contract NAS5-26555.

Simulations in this project and the mock catalogs generated from them were run on the Caltech compute cluster ``Wheeler,'' allocations from XSEDE TG-AST130039 and PRAC NSF.1713353 supported by the NSF, NASA HEC SMD-16-7592, and the High Performance Computing at Los Alamos National Lab., and analyzed using computing resources supported by the Scientific Computing Core at the Flatiron Institute.  Further simulations were run using Early Science Allocation 1923870 on Frontera, which is made possible by National Science Foundation award OAC-1818253. This work used additional computational resources of the University of Texas at Austin and TACC, the NASA Advanced Supercomputing (NAS) Division and the NASA Center for Climate Simulation (NCCS), and the Extreme Science and Engineering Discovery Environment (XSEDE), which is supported by National Science Foundation grant number OCI-1053575. Last but not the least, huge thanks to Jason Hunt who generously hosted the data as we were closing in on the finish line.

\appendix

\section{Yield prescriptions in the simulation}
\label{sec:appendix_yields}
As discussed in Section \ref{sec:sim_data}, the stellar evolution in FIRE-2 \citet{hopkins08} is produced from STARBURST99 \citep{leitherer99} with a \citet{kroupa01} IMF. The three sources of nucleosynthesis included are SNe II, SNIa, and stellar winds from OB/AGB stars. In the simulations, SNe II rates are a function of time (i.e. age of a star particle) dictated by the following: 
\begin{equation*}
\rm R_{SNeII}~(SNe~Myr^{-1}~M_{\odot}^{-1}) = \left\{
        \begin{array}{ll}
            0 & \quad \rm age < 3.4~Myr \\
            5.408 \times 10^{-4} & \quad \rm 3.4~Myr<age<10.37~Myr \\
            2.516 \times 10^{-4} & \quad \rm 10.37~Myr<age<37.53~Myr \\
            0 & \quad \rm age > 37.53~Myr,
        \end{array}
    \right.
\end{equation*}

with IMF-averaged ejecta mass of $\rm 10.5~M_{\odot}$ per explosion, most of which are in H. The SNe II yields for the elements in this study are listed on the first column of Table \ref{tab:yields}. On the other hand, SNIa rates have the following prescriptions: 
\begin{equation*}
\rm R_{SNIa}~(SNe~Myr^{-1}~M_{\odot}^{-1}) = \left\{
        \begin{array}{ll}
            0 & \quad \rm age < 37.53~Myr \\
            \rm 5.3 \times 10^{-8} + 1.6 \times 10^{-5} e^{-((age-50)/10)^{2}/2} & \quad \rm age > 37.53~Myr,
        \end{array}
    \right.
\end{equation*}
with ejecta mass of $\rm 1.4~M_{\odot}$ per SNIa explosion. The element yields per SNIa explosion are tabulated in the second column of Table \ref{tab:yields}. Lastly, mass-loss from OB/AGB stellar winds were included whose yields are listed on the third column of Table \ref{tab:yields}. For more details on the stellar evolution in FIRE-2, we refer to Appendix A in \citet{hopkins18}.

\begin{table}[!htbp]
\begin{center}
\caption{\textit{Element yields from different sources.} SNe II yields are from \citet{nomoto06}, SNIa from \citet{iwamoto99}, and stellar winds from the compilation by \citet{wiersma09} of \citet{vdhoek97}, \citet{marigo01}, and \citet{izzard04}. Yields are in terms of $\rm M_{\odot}$. We note that the N abundance has a metallicity dependence and is multiplied by \textit{N} where \textit{N}=max($Z/Z_{\odot}$,1.65).}
\begin{tabular}{cccc}
\hline \hline
Elements & SNe II & SNIa & OB/AGB winds \\
\hline
Fe & 0.0741 & 0.743 & -\\
C & 0.133 & 0.049 & 0.016 \\
N & 0.0479\textit{N} & 1.2 $\times 10^{-6}$  & 0.0041\\
O & 1.17 & 0.143 & 0.0118\textit{N} \\
Mg & 0.0987 & 0.0086 & -\\
Si & 0.0933 & 0.156 & -\\
S & 0.0397 & 0.087 & -\\
Ca & 0.00458 & 0.012 & -\\
\hline\hline
\end{tabular}
\label{tab:yields}
\end{center}
\end{table}

\section{Optimal age-binning}
\label{sec:appendix_bin}

In Section \ref{sec:kdes}, we determined the optimal binning in age to look at the $\rm R_{birth}$, current R, [Fe/H] and [X/Fe] distributions as a function of age. We started with a bin size of $\pm$ 0.5 Gyr and scrutinized the [Fe/H] vs $\rm R_{birth}$ of the stars encompassed by this selection. We reduce the bin size by half, and calculate how the dispersion across the linear fit changes. A decrease in dispersion warrants a further decrease in the age bin; an increase in dispersion or a sample size $<$50,000 stars stops the iteration. In general, we define the age bin as sufficient when it is (1) not too wide such that the measured dispersion in the [Fe/H] vs $\rm R_{birth}$ trend is independent of the binning and (2) not too narrow to be affected by small number statistics.

\begin{figure}
\centering
\includegraphics[width=0.45\textwidth]{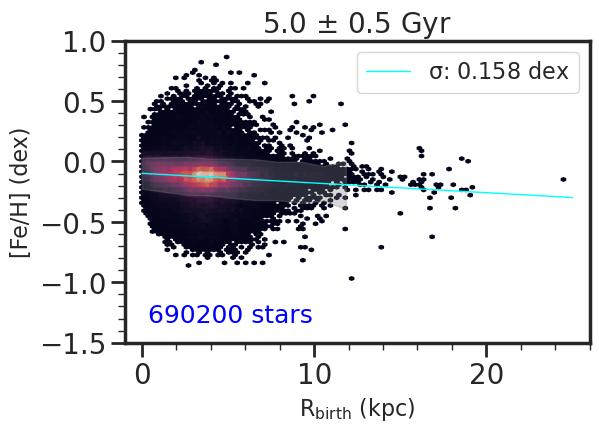} 
\includegraphics[width=0.45\textwidth]{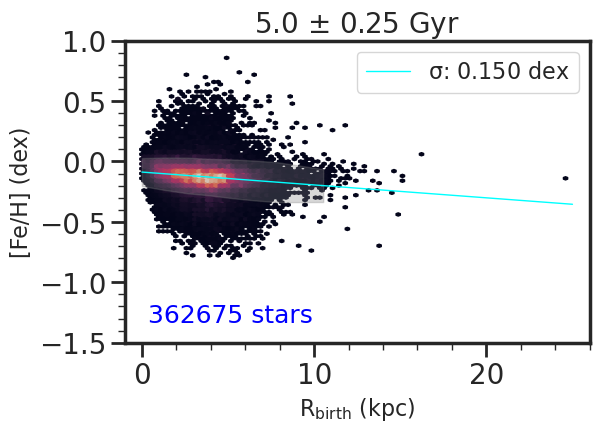} 
\includegraphics[width=0.45\textwidth]{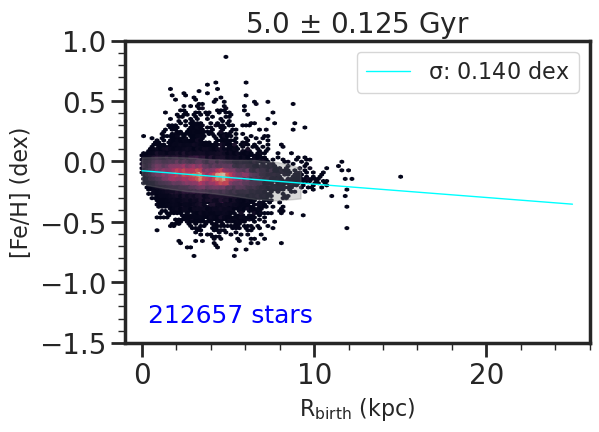} 
\includegraphics[width=0.45\textwidth]{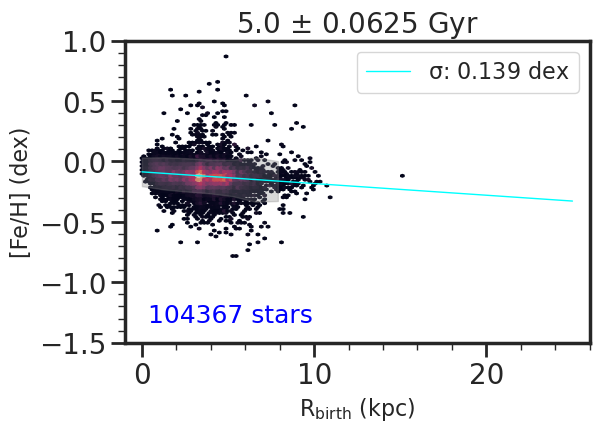} 
\includegraphics[width=0.45\textwidth]{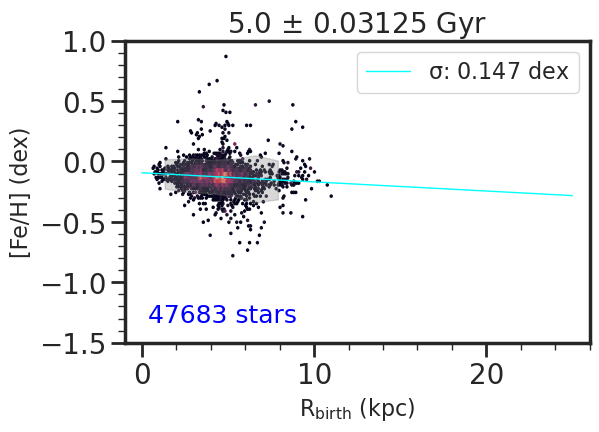} 
\caption{\textit{[Fe/H] vs $R_{birth}$ for different age binning.} Each panel shows this trend for different age bins of 5 Gyr old stars. We mark the best fit line (cyan) and the running dispersion around this trend (gray shaded region), and also note the number of stars included in the selection in the bottom left corner and the median of the running dispersion on the top right corner.} 
\label{fig:agebinning}
\end{figure}

Figure \ref{fig:agebinning} illustrates this process for 5 Gyr old stars, wherein we show their 2D histogram, with the best fit line for [Fe/H] vs $\rm R_{birth}$, and the running dispersion around this trend. We also note the number of stars included in the bin in the bottom left corner and the median running dispersion in the top right corner.

As we reduced the age bin size from 0.5 Gyr, the dispersion in [Fe/H] vs $\rm R_{birth}$ also continue to decrease until we reach 0.03 Gyr. At this point, the nth iteration has a small sample size that drives up the dispersion (e.g. 0.147 dex). We therefore choose the age binning from the n-1 iteration which gives us the smallest dispersion in the [Fe/H] vs $\rm R_{birth}$ diagram. We perform a similar exercise for the other ages included in the analysis. 

\bibliography{aamt}
\end{document}